\documentclass[a4paper,prl, superscriptaddress, reprint,longbibliography]{revtex4-2}

\def\Main{1}

\def\SI{1}

\usepackage{bibunits}
\usepackage[a4paper, margin=2.5cm]{geometry}
\usepackage{amsmath, amssymb, physics, bm}
\usepackage{graphicx}
\usepackage[dvipsnames]{xcolor}
\usepackage{hyperref}
\usepackage{cases}

\usepackage{braket}

\hypersetup{
    colorlinks=true,      
    linkcolor=blue,       
    citecolor=blue,      
    urlcolor=blue      
}

\newcommand{\tensorarrow}[1]{\overset{\text{\tiny$\bm\leftrightarrow$}}{\bm{#1}}}
\newcommand{\bvec}[1]{\ensuremath{\bm{#1}}}
\usepackage{esint} 
\renewcommand{\tensorarrow}[1]{\overset{\text{\tiny$\bm\leftrightarrow$}}{\bm{#1}}}
\usepackage{bm} 
\renewcommand{\vec}{\ensuremath{\bm}}

\renewcommand{\Im}{\text{Im}}
\renewcommand{\Re}{\text{Re}}
\newcommand{\multicomment}[1]{}
\defaultbibliographystyle{apsrev4-2}
\usepackage{microtype}  
\usepackage{silence}
\usepackage[utf8]{inputenc}
\usepackage[english]{babel}
     
\usepackage{graphicx}
\usepackage{amsmath,amssymb}
\usepackage{amsfonts}
\usepackage{mathtools}
\usepackage{xcolor}

\usepackage{braket}
\renewcommand{\eqref}[1]{(\ref{#1})}

\usepackage{titlesec}
\titleformat*{\section}{\raggedright\bfseries\sffamily\large}
\titleformat*{\subsection}{\raggedright\bfseries\sffamily}
\titleformat*{\subsubsection}{\raggedright\bfseries\sffamily}

\makeatletter
\renewcommand{\fnum@figure}{\textbf{Fig.~\thefigure}}
\makeatother
\usepackage{caption}
\usepackage{bm}
\renewcommand{\vec}{\ensuremath{\bm}}
\renewcommand{\Im}{\text{Im}}
\renewcommand{\Re}{\text{Re}}
\definecolor{bleu}{RGB}{0,47,108}
\usepackage{hyperref}
\hypersetup{colorlinks,citecolor={blue},linkcolor={red},urlcolor={red}}

\renewcommand{\eqref}[1]{(\ref{#1})}
\usepackage{enumitem}
\usepackage{pgfornament}
\usepackage{empheq}

\usepackage{xargs} 
\usepackage[colorinlistoftodos,prependcaption,textsize=footnotesize]{todonotes}
\newcommandx{\unsure}[2][1=]{\todo[linecolor=red,backgroundcolor=red!25,bordercolor=red,#1]{#2}}
\newcommandx{\change}[2][1=]{\todo[linecolor=blue,backgroundcolor=blue!25,bordercolor=blue,#1]{#2}}
\newcommandx{\info}[2][1=]{\todo[linecolor=OliveGreen,backgroundcolor=OliveGreen!25,bordercolor=OliveGreen,#1]{#2}}
\newcommandx{\improvement}[2][1=]{\todo[linecolor=Plum,backgroundcolor=Plum!25,bordercolor=Plum,#1]{#2}}
\newcommandx{\thiswillnotshow}[2][1=]{\todo[disable,#1]{#2}}

\DeclareMathOperator{\sgn}{sgn} 

\usepackage{nicematrix}

\usepackage{esint} 

\newcommand{\PRLsep}{\noindent\makebox[\linewidth]{\resizebox{0.3333\linewidth}{1pt}{$\bullet$}}\bigskip}

\newcommand{\Hel}{\mathcal{H}^\text{pl}}
\newcommand{\Held}{\mathcal{H}^{\text{pl,d}}}
\newcommand{\Heldnorm}{\overline{\mathcal{H}}^\text{pl,d}}

\newcommand{\gsca}{\ensuremath{g}}
\newcommand{\gscav}[1]{\ensuremath{g_{#1}}}

\begin{document}

\if\Main1
\begin{bibunit}

\title{\sffamily\Large Absolute Experimental Measurement of Intrinsic Helicity in Elementary Plasmonic Nanostructures}
\date{\today}

\author{Malo Bézard}
\affiliation{%
Universit\'e Paris-Saclay, CNRS, Laboratoire de Physique des Solides, 91405, Orsay, France
}%

\author{Simon Garrigou}
 \affiliation{Univ Toulouse, CNRS, CEMES, Toulouse, France}%

 \author{Jérémie Béal}
\affiliation{%
Lumière, nanomatériaux, nanotechnologies (L2n), UMR CNRS 7076, Université de Technologie de Troyes, Troyes 10004, France
}%

 \author{Andreas Horrer}
\affiliation{%
Lumière, nanomatériaux, nanotechnologies (L2n), UMR CNRS 7076, Université de Technologie de Troyes, Troyes 10004, France
}%

\author{Yves Auad}
\affiliation{%
 Universit\'e Paris-Saclay, CNRS, Laboratoire de Physique des Solides, 91405, Orsay, France
}%

\author{Hugo Lourenço-Martins}
\email{hugo.lourenco-martins@cnrs.fr}
 \affiliation{Univ Toulouse, CNRS, CEMES, Toulouse, France}%

\author{Davy Gérard}
 \email{davy.gerard@utt.fr}
\affiliation{%
Lumière, nanomatériaux, nanotechnologies (L2n), UMR CNRS 7076, Université de Technologie de Troyes, Troyes 10004, France
}%

\author{Mathieu Kociak}
 \email{mathieu.kociak@universite-paris-saclay.fr}
\affiliation{%
 Universit\'e Paris-Saclay, CNRS, Laboratoire de Physique des Solides, 91405, Orsay, France
}%

\date{\today}
\begin{abstract}
Helicity is a universal quantity describing the internal rotation of an object or a field. Whether it characterizes fundamental properties such as the spin of elementary particles or leads to practical consequences such as the  toxicity or harmlessness of chiral molecules, defining and measuring it is essential. However, this remains ambiguous in the case of chiral photonic systems such as plasmonic nanoparticles or photonic metasurfaces, where contrary to common knowledge, the observation of circularly polarized dependent optical properties is not a relevant measure for the helicity. We demonstrate experimentally and theoretically that the helicity can be rigorously defined and measured in a nanoplasmonics system emitting circularly polarized light after excitation in the near-field by a focused electron beam. In the case of a model system composed of two plasmonic dipoles (Born-Kuhn systems), we show that the helicity of photonic modes takes on a very intuitive form and can be simply measured  by symmetrizing the geometry of excitation and detection. The method could be extended to a variety of chiral photonic systems, whose locally enhanced properties make them promising for the engineering of local chirality.
\end{abstract}

\maketitle

\maketitle

Chirality, defined as the property of an object that cannot be superimposed on its mirror image, is one of the most pervasive asymmetries in nature, from elementary particles to biological macromolecules such as amino acids, sugars, and DNA \cite{wagniere2007chirality}. Crucially, chirality is not confined to matter: light itself can be chiral, as circularly polarized photons carry a well-defined helicity of $\pm 1$. Chiroptical interactions therefore sit at the heart of light-matter interactions in chiral systems, with far-reaching consequences across physics, chemistry, and biology, from pharmacology \cite{Contemporary_chiral_drugs_2024} and spintronics \cite{nvemec2018antiferromagnetic,Zutic2004,Physics_Report_YU_2023} to quantum information \cite{lodahl2017chiral}.

Chiroptical spectroscopies exploit this interaction to interrogate chiral properties through well-established linear techniques such as circular dichroism (CD), circularly polarized luminescence (CPL), infrared vibrational optical activity, and photoelectron circular dichroism, among others (see \cite{Berova2012} for a comprehensive review). These spectroscopies are routinely employed to analyze molecules in solution, where the associated physical observables are unambiguously defined. For example, CD corresponds to the differential absorption of left- and right-handed circularly polarized light by chiral molecules in solution and is regarded as an unequivocal signature of molecular chirality. 

    Beyond molecules in solution, the rapid development of nanostructures engineered to tailor the chirality of electromagnetic fields in both the near and far field \cite{Collins2017} has opened an entirely new regime, in which chiroptical coupling can be interrogated and engineered at the nanoscale. New characterization techniques have been proposed over the past two decades to meet this challenge \cite{Kwon2023,buchner2025wide}, including nonlinear chiroptical approaches based on multiphoton processes, such as second-harmonic generation circular dichroism and optical rotation \cite{Collins2018}, and, most recently, hyper-Raman optical activity \cite{jones2024chirality}. Some of these techniques aim to extend established chiroptical metrics to the nanoscale, such as circular differential scattering \cite{Wang2015,Thomas2026}, which generalizes CD to plasmonic nanostructures (hereafter referred to as circular dichroism in scattering, CDS). Others have introduced conceptually new quantities, particularly in attempts to quantify what is often loosely termed "near-field chirality", that is, the local chirality of the electromagnetic field. While a rigorous definition of optical chirality, directly connected to the enantioselective excitation rates of chiral molecules, has been established \cite{tang_optical_2010-1}, several studies have focused on experimentally accessible approaches that merely characterize the dissymmetry of the electromagnetic near field \cite{Hashiyada2014,Horrer2020,Aoudjit2023}, or have drawn connections with the radiative local density of states \cite{Zu2018,Pham2018}. This diversity of definitions has led to substantial confusion in the literature regarding the notion of optical chirality in nanostructures. Further compounding this confusion, CDS can be detected even on 2D achiral nanostructures, whether in the near-field \cite{Guido2025,Wang2015,Hashiyada2014,Osorio2016,Zu2018,Oshikiri2021,Sannomiya2021}, or far-field \cite{Mildner2023}. This is best exemplified by polarized cathodoluminescence (pCL) \cite{Zu2018,Osorio2016,Sannomiya2021}, a technique that has consistently revealed strong variations of CDS contrast at the nanometre scale in both achiral \cite{Zu2018,Sannomiya2021} and chiral \cite{Fang2016,Lingstadt2023} structures.

The contrast between the well-defined chiroptical observables of molecules in solution and the ambiguous metrics used to characterize chiral photonic excitations in nanostructures represents a significant challenge for the field. A clear definition and a reliable experimental quantification of the helicity of plasmonic excitations are therefore needed.

 Here we propose a definition of helicity for plasmonic structures and an experimental method to measure it. We explicitly demonstrate its applicability to the most elementary chiral plasmonic system, composed of two offset coupled dipoles exhibiting a bonding and an antibonding mode of opposite chirality. Its conceptual simplicity and the availability of analytical expressions for the plasmonic modes' helicity allow us to confirm all the fundamental chirality-related symmetries revealed by first-principles calculations. Specifically, we consider two coupled dipoles, experimentally realized as two plasmonic gold antennas (Born-Kuhn systems, BKS), excited by a focused free-electron beam and generating a circularly polarized pCL signal. We show that proper symmetrization of the pCL excitation and detection geometries leads to the direct experimental measurement of the modes' helicity. These definitions and methodologies are parsimonious and will provide a foundation for future studies of chiral properties in polaritonic systems well beyond plasmonics.

\section{Results}
 \subsection{Theory}
 \subsubsection{Fundamental aspects of plasmonic chirality}

The difficulties outlined above are largely conceptual and stem from the ambiguity of the term \emph{chirality} in nanophotonics, best illustrated in the context of plasmonics. For example, "chirality" may refer to the geometrical chirality of nanostructures sustaining nano-optical excitations, quantified by the Kelvin chirality $\mathcal{K}$ \cite{nechayev_kelvins_2021}.

In this sense, the Kelvin chirality is the property of an object whose symmetry group contains no parity transformation (inversion, mirror, or improper rotation).  $\mathcal{K}=0$ denotes an achiral structure, while $\mathcal{K} =\pm 1$ refers to left- and right-handed chiral structures that are mirror-symmetric (i.e., interconverted by the parity operation $\vec{r} \rightarrow -\vec{r}$). But if one refers to the \textit{local} chirality of the electromagnetic (EM) field, one needs to use the celebrated density of optical chirality usually noted $\mathcal{C}$ \cite{tang_optical_2010-1}. Finally, optical excitations in matter are polaritonic in essence. Therefore, their potential chiral symmetries should be associated with  another  quantity, yet to be defined, that  accounts for the polarization of the medium. We denote this quantity $\Hel$ for reasons that will become clear later.  One can therefore see that, because it involves several distinct notions $\mathcal{K}, \Hel, \mathcal{C}$, the hybrid light-matter and local nature of polaritons are the main source of difficulties with the characterization of their chirality and its relation to the CDS, should it be local or global. This question of defining the optical chirality in complex dispersive and dissipative media is currently under intense theoretical investigation \cite{alpeggiani_electromagnetic_2018, vazquez-lozano_optical_2018, fernandez-corbaton_total_2021,moser2026unifiedlightmattermetricmolecular}. In particular, recent studies \cite{mackinnon_quantized_2025} employing the Hopfield model have demonstrated that part of the helicity of a polariton is stored by the matter field. In this work, we  focus on one simple example of polaritons - localized surface plasmons (LSP) - and show that their helicity is fully determined by their matter degree of freedom i.e., the polarization field $\vec{\mathfrak{p}}(\vec{r},t)$. This is expected, since an LSP is a lower-branch polariton in the long-wavevector limit, where the excitation acquires a predominantly matter-like character.\\ 

This problem can be tackled with a general and powerful notion of vector analysis: the helicity  $\mathcal{H}=\hat{\mathcal{H}}[\vec{u}]= \vec{u}.\vec{\nabla}\times\vec{u}$ of a 3-vector field $\vec{u}(\vec{r},t)$ \cite{efrati_orientation-dependent_2014}. For example, in hydrodynamics, as applied to a fluid velocity field $\vec{v}$, $\hat{\mathcal{H}}[\vec{v}]$ characterizes the knottedness of a vortex flow. Remarkably, the free-space optical chirality $\mathcal{C}$ is simply the sum of the electric and magnetic helicity $\mathcal{C}(\vec{r},t)=1/2(\hat{\mathcal{H}}[\vec{E}(\vec{r},t)]+\hat{\mathcal{H}}[\vec{B}(\vec{r},t)])$ \cite{vazquez-lozano_optical_2018}. Due to the dual symmetry (i.e., the internal rotation of the EM field) of the free Maxwell's equations \cite{berry_optical_2009}, the Noether theorem imposes the conservation of chirality \cite{cameron_electricmagnetic_2012,bliokh_dual_2013}, i.e. $\partial_t \mathcal{C}+\vec{\nabla}.\vec{\mathcal{S}}=0$ with $\vec{\mathcal{S}}$ the EM spin current. 
This implies that the total chirality variation inside the volume $\mathcal{C}_{tot}=\iiint_{\Omega} \mathcal{C} d^3\vec{r}$ encompassed in the far-field sphere is directly proportional to the spin flux $\vec{\mathcal{S}}$ through its surface $\partial V$, $\oiint_{\partial V} \vec{\mathcal{S}}.d\vec{s}$ \cite{poulikakos_optical_2016}. Noting $I_R (\vec{k})$ ($I_L(\vec{k})$) the right- (left-) handed circularly polarized light intensity along direction $\vec{k}/k$,  the latter is simply the  Stokes vector component $S_3 = I_R (\vec{k}) -I_L (\vec{k})$, integrated over $\partial V$ i.e. the total CDS. Following the convention in the literature, we will use a quantity normalized by the total (left-handed plus right-handed circularly polarized, i.e,  the zeroth-component of the Stokes vector $S_0$) intensity integrated on the detection angle. This quantity is known as the  dissymmetry factor $\gsca$  \cite{Guido2025}. Measured on a given solid angle $\Omega$, it reads $g_{\Omega}= 2\frac{\int\int_{\Omega}d\Omega S_3}{\int\int_{\Omega} d\Omega S_0}$. Thus, in free-space and for a monochromatic field of frequency $\omega$, the total CDS ~\gscav{4 \pi} ~  is a direct measurement of the total chirality of the near-field, i.e:
\begin{equation}
    \gscav{4 \pi}\propto\mathcal{C}_\text{tot}(\omega)
    \label{eq:Ctot}
\end{equation}

In other words, while the local helicity of the plasmonic EM field lines, encoded in the space-dependent $\mathcal{C}(\vec{r},\omega) \propto \Im\{ \vec{E}.\vec{B}^* \}$ \cite{schaferling_tailoring_2012}, does not uniquely determine the measured CDS, its total integrated value $\mathcal{C}_\text{tot}$ does.   

To avoid considering the difficulties associated with dealing with complex polaritonic media, one can concentrate on plasmonic systems, calculating the plasmonic modes $m$ \cite{Boudarham2012} and their associated total chiralities $\mathcal{C}_{\text{tot},m}$, each contributing separately to $\gsca(\omega)$ through its associated time-dependent $\vec{\mathfrak{p}}_m(t)$ or energy-dependent $\vec{p}_m(\omega)$ eigendipole density. Following this approach, we can then directly introduce the \textit{modal plasmonic helicity} for mode $m$ as:

\begin{flalign}
\label{EQ:helicity_polariton}
\Hel_m (t) &= \int \hat{\mathcal{H}}[\vec{\mathfrak{p}}_m(\vec{r},t)] d^3\vec{r}&& \\   &= \int \vec{\mathfrak{p}}_m(\vec{r},t).\vec{\nabla}\times\vec{\mathfrak{p}}_m(\vec{r},t) d^3\vec{r} 
\end{flalign}

\noindent or equivalently in the energy domain:
\begin{equation}
    \Hel_m (\omega)= \frac{1}{2} \Re{ \int_{\mathbb{R}^3} [\vec{p}^*_m(\vec{r},\omega).\vec{\nabla}\times\vec{p}_m(\vec{r},\omega)] d^3\vec{r}}
    \label{EQ:helicity_polariton_cmplx} 
\end{equation}

\noindent The total helicity is then the sum of the helicity of each mode. While the equivalence between $\mathcal{C}_{\text{tot}}$ and $\gscav{4\pi}$ (Eq. \ref{eq:Ctot}) is valid for arbitrary nano-optical systems, we will show in this work that for elementary plasmonic systems, this equality can further be reduced to
\begin{equation}
\Hel_m (\omega) \propto \gscav{4\pi}
\label{eq:hel_is_g}
\end{equation}

 We therefore argue that equations \eqref{EQ:helicity_polariton} to \eqref{EQ:helicity_polariton_cmplx} form the most appropriate definition of the plasmonic chirality, which can be directly related to the CDS when the proper experimental geometry is defined through equation \eqref{eq:hel_is_g}. First, if the nano-structure is geometrically achiral (chiral) $\mathcal{K}=0$ ($\mathcal{K}\neq0$), then the dipole density $\vec{p}_m(\vec{r},\omega)$ must (not) be parity symmetric and therefore $\Hel_m \equiv 0$ ($\Hel_m \not\equiv 0$) and $g(\omega)=0$ ($g(\omega)\neq 0$). There is therefore a direct connection between $\mathcal{K}$ and the circular dichroism scattering signal integrated on $4\pi$, $\gscav{4\pi}(\omega)$. This resolves the apparent conceptual tension raised in the introduction. Second, our definition satisfies a principle of parsimony: it does not require the addition of any quantity, and only involves the helicity operator, as for any field. 
Thus, we have seen that two fundamental ingredients - the dual symmetry and the helicity operator - are sufficient to give a complete picture of the plasmonic chirality and its relation to CDS. In the next section, we will see that our result can be connected and reduced to other theoretical works in the literature, when applied on a simple toy model: the Born-Kuhn plasmonic system (BKS) \cite{Schaferling2017}.

 \subsubsection{Application to a toy system from first principles: the BKS}

\begin{figure*}
    \centering
    \includegraphics[width=1.0\linewidth]{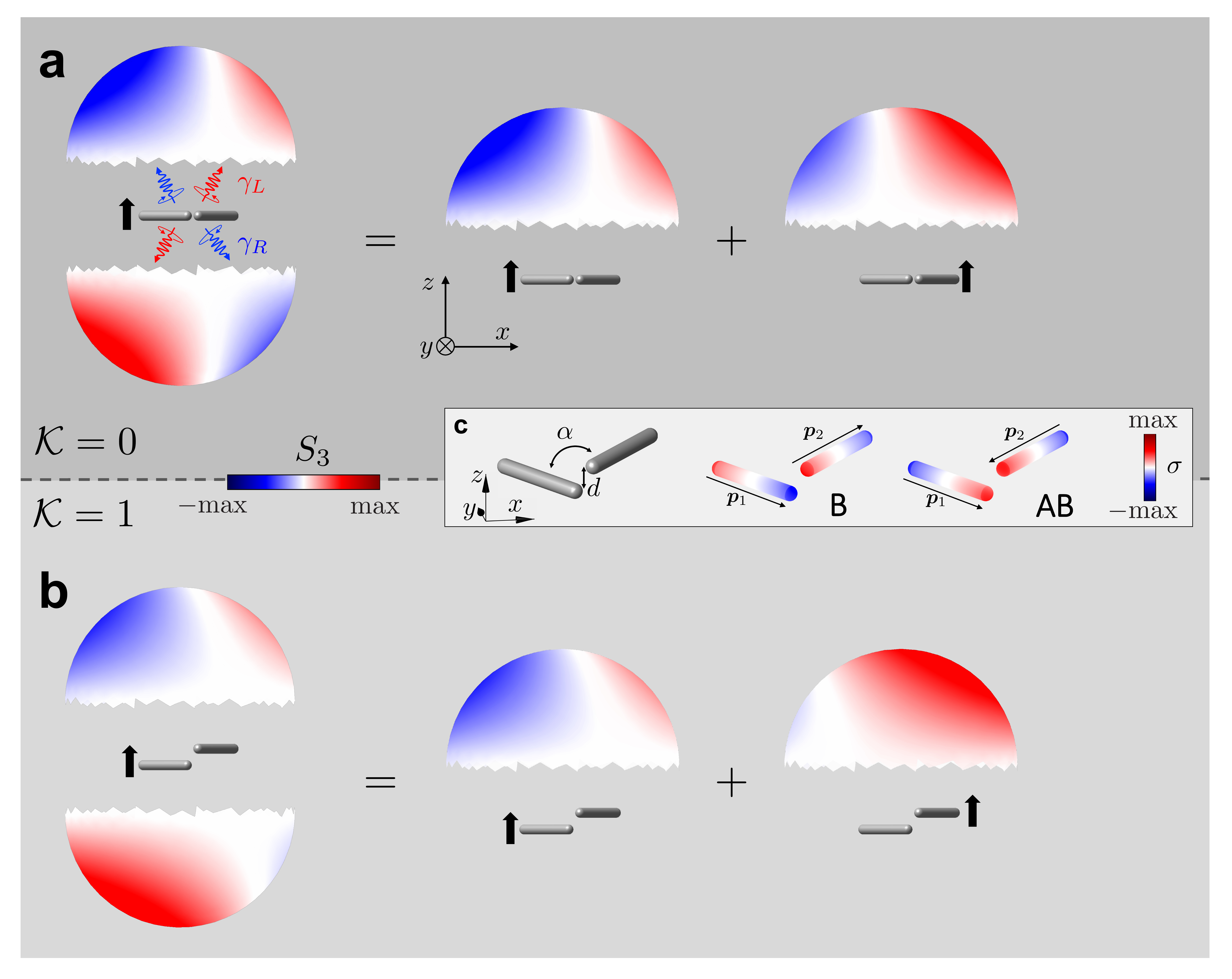}
    \caption{\textbf{Principle of the measurement of helicity $\Hel$}. \textbf{(a)} an achiral $\mathcal{K}=0$ or \textbf{(b)} a chiral $\mathcal{K}=1$ system (here, a plasmonic BKS structure, see \textbf{(c)}) is excited locally with an electron beam represented by a black arrow. Its angular-dependent CDS circular dichroic emission $S_3$ is collected over the whole far-field sphere, and $\Hel$ is equal to its integral. $S_3$ is represented on the north and south hemispheres, as viewed along the $y$ axis (see c). The schematic is not to scale.  The clear mirror parity symmetry between the north and south hemispheres in the achiral case leads to a null integrated CDS and $\Hel=0$. This symmetry is absent for the chiral structure, thus leading to non-null integrated CDS and $\Hel\neq 0$. Such a geometry is experimentally intractable. It can be mimicked by performing two symmetric, experimentally tractable, circularly polarized experiments on the north hemisphere only, whose results are summed. $S_3$ far-field signal is here simulated for a BKS bonding mode (see \textbf{(c)}) with the \texttt{MNPBEM} \cite{Hohenester2014} toolbox. \textbf{(c)} left: schematic and parameters for a BKS system. right: eigencharge $\sigma$ distribution for the bonding (B) and antibonding (AB) mode simulated with \texttt{MNPBEM}. \cite{Hohenester2014}}
    \label{fig:principles}
\end{figure*}

 The BKS system excited by an electrical dipole, shown for an achiral ($\mathcal{K}=0$, Fig.~\ref{fig:principles}a) and chiral ($\mathcal{K}=+1$ in Fig.~\ref{fig:principles}b) configuration, is made up of two coupled plasmonic dipoles (see the corresponding charge distribution on Fig.~\ref{fig:principles}c). It has been amply studied theoretically in the context of electron-based spectroscopies \cite{Lourenco-Martins2021,Bourgeois2022,Konecna2023} and  could be realized experimentally with two shifted gold nano-antennas placed in two parallel planes separated by a gap $d$ and forming an angle $\alpha$ (see inset of Fig.~\ref{fig:principles}(c)). The uncoupled dipole energy is $\omega_0$ and the coupling (with coupling constant $\omega_c$) leads to the emergence of a bonding (energy $\omega_-$) and anti-bonding ($\omega_+$) modes. The bonding (B) and anti-bonding modes (AB) eigencharges are shown in Fig.~\ref{fig:principles}(c)). This idealized system makes it easy to unravel the link between local CDS and chirality. Here, we reduce its polarization density to (see SI, section S1 B):
\begin{equation}
  \vec{\mathfrak{p}}(\vec{r})=\underbrace{\vec{\mathfrak{p}}_1 \; f_\varepsilon (\vec{r}-\vec{r}_1)}_{=\mathfrak{p}_1(\vec{r})}+\underbrace{\vec{\mathfrak{p}}_2 \; f_\varepsilon(\vec{r}-\vec{r}_2)}_{=\mathfrak{p}_2(\vec{r})}
  \label{EQ:smoothed_plasmon_field}
\end{equation}
\noindent where $f_\varepsilon$ is a regularization function.

Then, by applying the definition \eqref{EQ:helicity_polariton} (see SI section S1) one can show that the plasmonic helicity of the BKS system made of two dipoles $\vec{\mathfrak{p}}_{1,2}$ is given by  
 \begin{equation}
    \mathcal{H}^\text{pl} = \aleph \, (\vec{\mathfrak{p}}_1\times \vec{\mathfrak{p}}_2).(\vec{r}_1-\vec{r}_2) \propto \sgn(\vec{\mathfrak{p}}_1 \cdot \vec{\mathfrak{p}}_2) \cdot d\cdot \sin\alpha
    \label{EQ:Helicity_BKS}
 \end{equation}
\noindent where $\aleph$ is a regularization constant, see (SI) and $\sgn$ is the sign function. Here, the subscripts in $\vec{\mathfrak{p}}_{1,2}$ refer to individual dipoles constituting the BKS, not to the coupled dipoles that write $\vec{\mathfrak{p}}_{B,AB}=\vec{\mathfrak{p}}_{1}\mp\vec{\mathfrak{p}}_{2}$ for the bonding (B) and antibonding (AB) configurations. Remarkably, this expression provides the \emph{explicit} dependence of plasmonic chirality on Kelvin chirality of the nano-structures, showing that the former vanishes for a mirror-symmetric system i.e. when the gap $d=0$ or when the relative orientation of the two dipoles reaches $0$ or $180^\circ$. Additionally, it properly changes sign when the dipoles' relative polarization or structural handedness is flipped. Finally, it is important to emphasize that this formula  reproduces other definitions of BKS chirality recently given in the literature \cite{moser_conservation_2025,ossikovski_bornkuhn_2025}. In addition, treating the BKS system with a two-group model \cite{andrews_quantum_2018}, the present definition exactly matches the  general definition of dipole chirality $\Im\{\vec{p}.\vec{m}^*\}$, where $\vec{m}$ is the magnetic moment of the dipole system (see the SI section S1D for a detailed derivation).\\

 We next propose a measurement method for the plasmonic helicity. We present the distribution of $S_3$ over the far-field sphere of achiral and chiral BKS subject to a local excitation in Fig.~\ref{fig:principles}.  In the case of the achiral structure ($\mathcal{K}=0$, Fig.~\ref{fig:principles}a), $S_3$ has a perfect parity mirror symmetry. It follows that $\gscav{4\pi} =0 $ when the structure is excited at one of its tips. As shown in Fig.~\ref{fig:principles}(a-b) the symmetry is broken  for a chiral BKS ($\mathcal{K}=1$, Fig.~\ref{fig:principles}b) and $\gscav{4\pi} \neq 0 $. The simulations are here performed for the bonding mode and a $\mathcal{K}=+1$ structure, but the conclusion also holds for the antibonding mode and/or a $\mathcal{K}=-1$ structure. This confirms in the case of BKS that indeed $\Hel$ can be obtained by integrating the CDS signal over the full far-field sphere.  Unfortunately, it is experimentally impossible to perform a $4 \pi$ integration of the emission signal (which will be considered to be a pCL signal for the rest of this paper), as collection mirrors usually cover only one of the two hemispheres. If we were to measure only one of the two hemispheres ($\gscav{\pm 2\pi}$), the parity symmetry of the combination of the structure, the excitation and the detection is broken, possibly inducing \textit{extrinsic} chirality. It explains why CDS can emerge for \textit{achiral} structures \cite{Hashiyada2014,Osorio2016,Zu2018,Han2018RevealSubnanoscale,Horrer2020,Oshikiri2021,Sannomiya2021,Mildner2023} if the experimental setup is not parity-symmetric.\\

In the present case, Fig.~\ref{fig:principles}(a-b) provides a hint on how to symmetrize properly the experiments. Indeed, in the achiral case, changing the excitation from one tip (tip 1) to the other (tip 2) reverses the $S_3$ distribution,  i.e $S_3(1,\vec{k})=-S_3(2,-\vec{k})$. Taking advantage of this symmetry, we obtain:
\begin{align}
    \gscav{4\pi} &\equiv \gscav{4\pi}(1)=\gscav{4\pi}(2) \\
    &= \gscav{\pm2\pi}(1)+\gscav{\pm2\pi}(2)
\end{align}
In other words, \gscav{4\pi} and therefore $\Hel$ can be measured by summing the  integrated CDS on one hemisphere only for two symmetric positions of the electron beam (the two external tips of the BKS in this case).

\subsubsection{Application to the BKS system using a two coupled oscillating dipoles model}
In order to substantiate the above discussion, we first explicitly introduce the energy dependence through a model of two \textit{oscillating} (energy-dependent) dipoles. The time-averaged helicity is given by (see SI section S1):
\begin{equation} \label{EQ:harmonic_plasmon_helicity}
    \braket{\mathcal{H}^\text{pl}}_T = \frac{\aleph}{2} \, \Re \left\{\vec{p}_1(\omega)\times \vec{p}^*_2(\omega) \right\}. (\vec{r}_1-\vec{r}_2)
\end{equation}

\noindent with $\aleph$ being a regularization constant (see SI section S1. B4 ). We then introduce a model of two coupled dipoles for which the phase shift is directly taken into account, leading to the intensity-normalized helicity  $\Heldnorm$ (see SI section S1 C and S2 G):
\begin{equation}
 \Heldnorm(\alpha,\omega) \propto \frac{\omega_c^2(\omega^2_0- \omega_c^2)}{|\Omega^2|^2 +\omega_c^4} \cdot d \sin\alpha
 \label{eq:held}
\end{equation}
with $\Omega = \omega_0^2-\omega^2+i\gamma\omega$, $\gamma$ being a term describing the dissipation of the plasmon modes. $\Heldnorm(\alpha,\omega)$ possesses two optima at $\omega_\pm$ with increasing contrast as the dissipation goes to zero, i.e:
\begin{equation}
    \Heldnorm(\alpha,\omega) \xrightarrow[\gamma \to 0]{} \Heldnorm_{-}(\alpha)+\Heldnorm_{+}(\alpha)
\end{equation}
where $\Heldnorm_{-(+)}(\alpha)$ is the helicity at the energy of the B (AB) mode. In other words,, $\Heldnorm(\alpha)$ is the sum of the helicity of the B and AB modes. 

As expected for helicity, it continuously approaches zero as $d$ and $\alpha ~ mod(\pi)$. It changes sign at the resonance energy (wavelength) of a single, uncoupled antenna $\omega_0$ (wavelength $\lambda_0$), having optima of opposite signs at the B and AB energies (wavelengths) $\omega_\mp$ ($\lambda_\mp$). The sign of the optima changes with the handedness ($d\leftrightarrow -d$ or $\alpha \leftrightarrow \alpha+\pi$). As predicted earlier from first principles,  $\Heldnorm$ exhibits all the expected symmetries for the helicity of a plasmonic BKS. 

Then, we calculate the value of $S_3$ for the case where an electron hits one tip or the other  ($j= 1 ~or ~2$) of a BKS with angle $\alpha$. We normalize $S_3$ to the sum of the left- and right-polarized intensities integrated over the detection volume (see SI S2 D),  $\bar{S_3}(j)(\omega,\alpha)$. Considering normalized quantities will make comparison with experiments easier. At this point, $\bar{S_3}(j)$ is not a quantity integrated over the emission direction $\vec{k}/|\vec{k}|$ where  $\vec{k}$ is the wavevector of the outgoing wave, i.e., it is not \gsca. We expand it as a function of $\vec{k}.(\vec{r}_1-\vec{r}_2)$. $\vec{k}.(\vec{r}_1-\vec{r}_2)\ll1$ because the distance between the two antennas is much smaller than a wavelength of light.

The zero-order term reads:
\begin{equation}
    \bar{S_3}^{(0)}(j) \propto (-1)^j \gamma\omega_c^2 \cdot sin(\alpha)\cos(\theta)\ \cdot  \frac{ \omega}{|\Omega^2|^2+\omega_c^4}
    \label{eq:zeroorder}
\end{equation}
 where $\theta$ is the polar angle. As expected (see Figure \ref{fig:principles}a), it is antisymmetric with the polar angle ($\theta \leftrightarrow \theta + \pi$) and upon swapping the tips. In other words, it integrates to zero over the sphere, or when summed over the two tips, regardless of the solid angle considered (see SI S2 C and E ).
 
 It also vanishes in the absence of dissipation ($\gamma =0$), i.e when the B and AB modes do not overlap. It peaks at $\omega_0$ ($\lambda_0$), i.e, at the resonant energy (wavelength) of individual, uncoupled dipoles. This phenomenological model quantitatively explains  the "hidden chirality" effect (see \cite{Zu2018,Xie2025}), characterized by the emission of strongly circularly polarized light, with handedness varying rapidly at the scale of a nanoparticle, even if the nanoparticle is not chiral. It is understood as an \textit{extrinsic} chirality effect arising from the interference of two geometrically orthogonal plasmonic dipoles at their overlapping wavelength ($\lambda_0$), and enabled by a sizable dissipation ($\gamma \neq 0$). We show here that this mechanism applies to chiral systems as well. Already established as a dominant effect in achiral systems \cite{Zu2018,Horrer2020}, it also dominates the CDS signal  for chiral systems  in the absence of proper symmetrization of the experimental geometry, as it constitutes a zeroth-order effect.

The first-order term reads 
 \begin{flalign}
\bar{S_3}^{(1)} (j)&   
\propto \sin(\alpha)\cdot \cos(\theta) && \\
\nonumber&\cdot \frac{\omega_c^2 }{|\Omega^2|^2+\omega_c^4} \left[   (\omega_0^2-\omega^2) \mathbf{k} \cdot (\mathbf{r}_1-\mathbf{r}_2) \right]&&
\label{eq:dvplt}
\end{flalign}
 
Unlike the zero-order term, the prefactors do not sum to zero when integrating over $4\pi$ or $2\pi$. The total normalized CDS signal, summed over the two tips  and integrated over one hemisphere ($\gscav{\pm 2\pi}(1,\alpha,\omega)$ and $\gscav{\pm 2\pi}(2,\alpha,\omega)$), therefore reads (see SI S2 G):

\begin{align}
  \sum_j\gscav{(-1)^{j}2\pi}(j,\alpha,\omega) =\gscav{4\pi}(\alpha,\omega) \\
  \propto k\Heldnorm(\alpha,\omega)+O((kd)^2)  
\end{align}

This shows that, within the two-dipole model, the proportionality between \gscav{4\pi} and the helicity is recovered. This also demonstrates that the helicity can be measured from the pCL dichroic signal ($S_3$ integrated on $2\pi$), summed over two different tips of a BKS. The case of a more realistic mirror is described in sections 2E and  8 of the SI. Analysis shows that this conclusion holds for the zeroth-order term, which vanishes regardless of the mirror symmetry. In the case of a revolution-symmetric mirror,  the dependence on $\omega$, $\alpha$  and $d$ (i.e the chiral behaviour of the plasmon modes) is preserved, with the measured \gsca ~ being slightly reduced relative to \gscav{4\pi}. If the mirror is non-symmetric, such as in the case of a parabolic mirror, an additional $2\alpha$ dependence may be introduced. Simulations show that the asymmetric but high-numerical-aperture mirror used here averages out this effect over a relatively large range of orientations of the BKS with respect to the parabolic rotation axis (see SI S2 E).

\subsection{Experimental and simulation results}

\begin{figure}[h!]
    \centering
    \includegraphics[width=\linewidth]{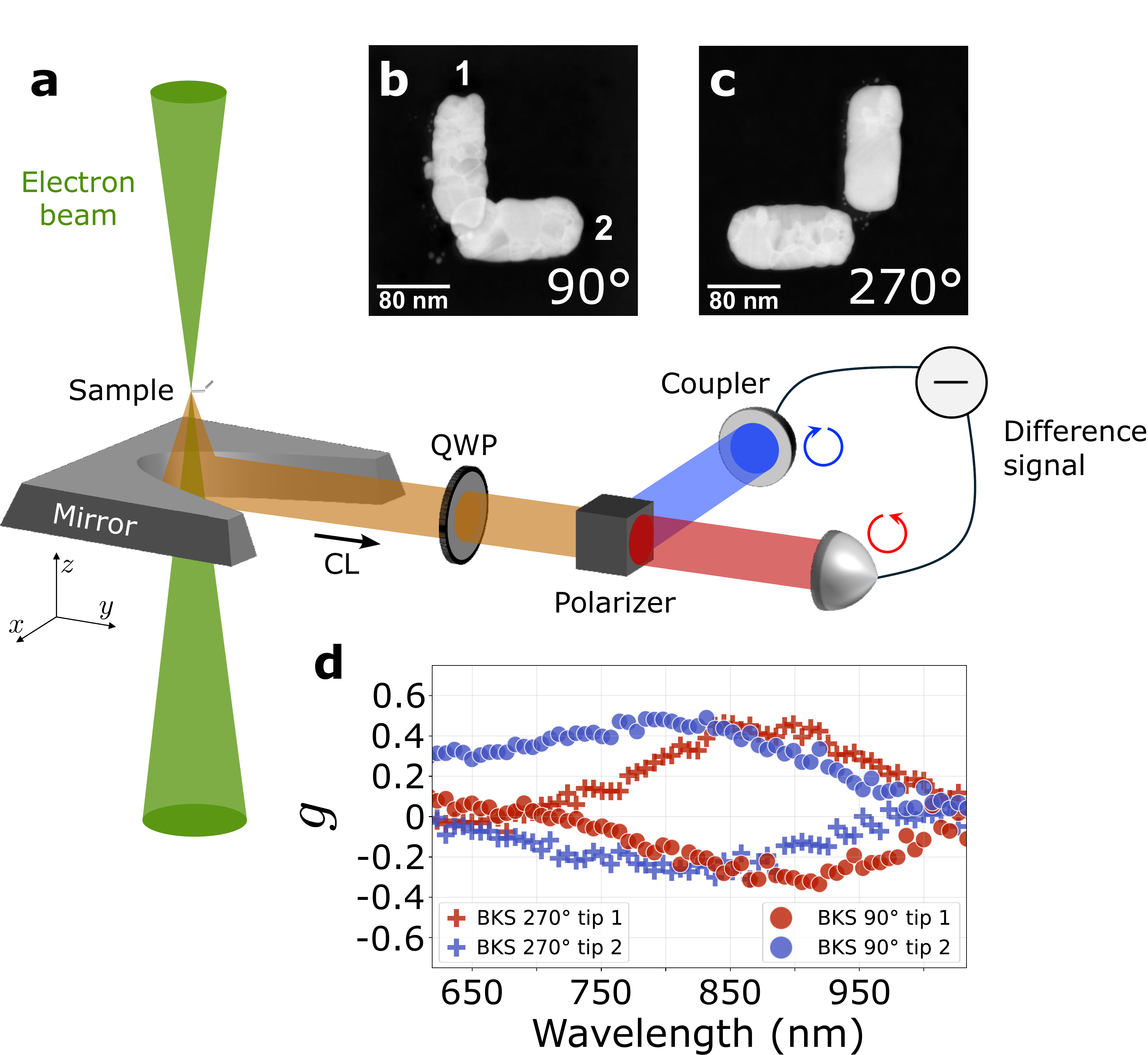}
    \caption{\textbf{BKS samples and STEM-pCL set-up.}  \textbf{(a)} The light emitted from a BKS sample under electron irradiation (cathodoluminescence), is collected by a parabolic mirror. HAADF images of a 90$^\circ$ and of a 270$^\circ$ BKS are shown in \textbf{(b)} and \textbf{(c)}. The two polarizations of the  collimated beam are filtered with a quarter-wavelength phase plate (QWP) and a linearly polarizing beam splitter (Polarizer) to be sent to two different optical fibers. The left and right polarized spectra in \textbf{(d)} are then measured simultaneously and their difference  ($g$) recorded. }
    \label{fig:setup}
\end{figure}

 pCL experiments were performed in a STEM using a custom-made drift-noise-free system (see Figure \ref{fig:setup}a), enabling simultaneous recording of the emission spectra for each electron beam position and for both circular polarizations, integrated over a parabolic mirror positioned in the bottom half-space of the sample. As described in the Methods and SI, this approach is essential for achieving the signal-to-noise ratio necessary to measure \gscav{\Omega_{exp}} over the experimental solid angle $\Omega_{exp}$. BKS samples were specially designed to be compatible with electron spectroscopy (see Methods and SI), and consist of two gold nano-antennas (150 nm in length, 50 nm in width, and 30 nm in thickness) — see the high-angle annular dark-field (HAADF) images of 90$^\circ$ and 270$^\circ$ BKS in Figure \ref{fig:setup}b and c. The distance between the top surface of the bottom antenna and the bottom surface of the top antenna is 10 nm. They exhibit strongly coupled B and AB modes (Rabi energy on the order of 350 meV \cite{Kociak2025}, with typical energies of 1.2 eV and 1.5 eV, respectively). Simulations were performed using MNPBEM \cite{Hohenester2014} and pyGDM \cite{wiecha2022pygdm}, without accounting for the substrate.
 \begin{figure}[h!]
    \centering
    \includegraphics[width=1.0\linewidth]{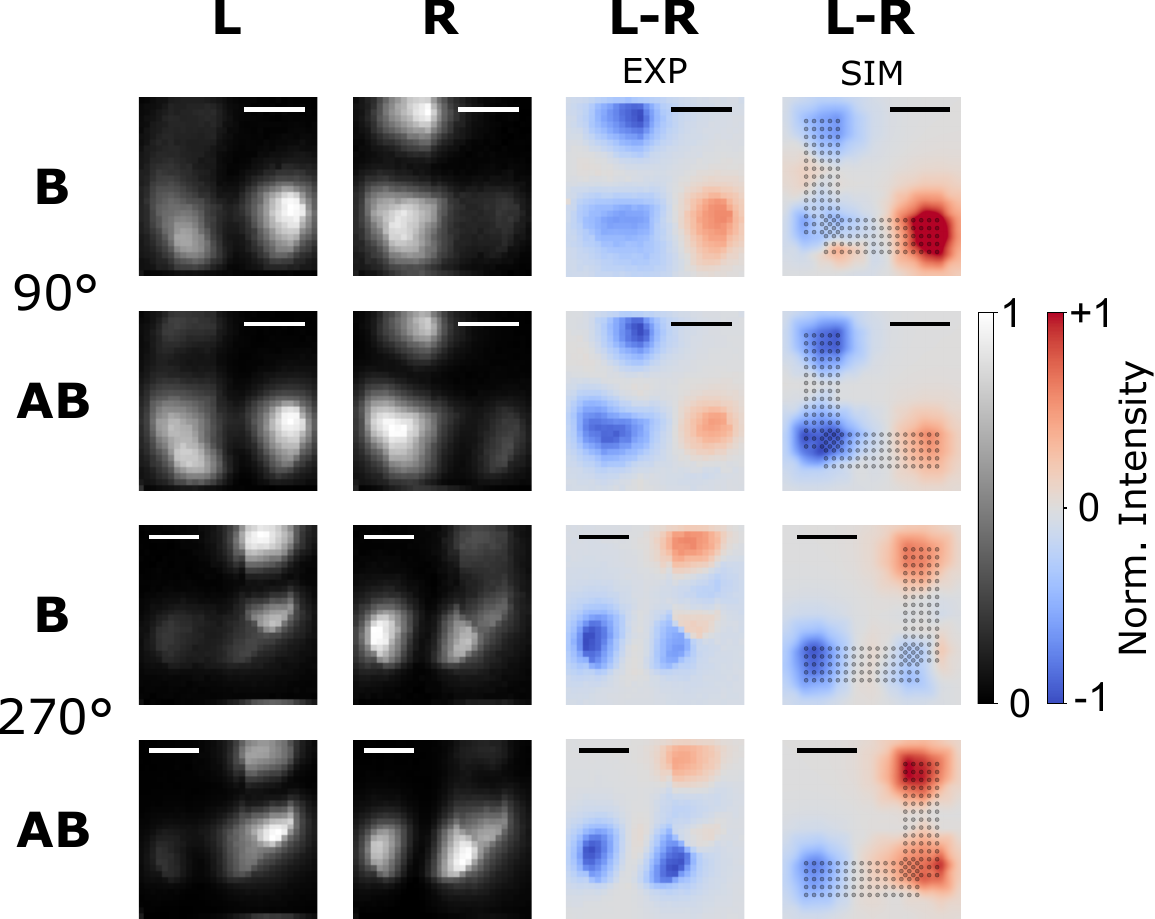}
    \caption{\textbf{Spatially localized circularly polarized CL}. \textbf{(a)} From pCL  spectral imaging, one can compute maps filtered at different energies, as exemplified in the specific case of a 90$^\circ$ BKS  and its handedness-symmetric counterpart, a 270$^\circ$ BKS.  The maps of the  left- and right-polarized CL (pCL L and pCL R) and of  their normalized difference (\gsca)  are displayed. The signal is integrated on the whole mirror collection angle and is summed on the whole range of the considered energy windows, bonding mode (B) region (850 - 940 nm) and antibonding mode (AB) region (760 - 830 nm). The right column represents the corresponding pyGDM simulated dichroic maps filtered at 898 nm and 738 nm for the B and AB modes respectively. Each scale bar represents 75 nm.  }
    \label{fig:chiral_map}
\end{figure}

 We now present a summary of the experimental and simulation results obtained on BKS with 12 different angles. Figure \ref{fig:chiral_map}a shows representative pCL experimental measurements for a 90° BKS and its handedness-symmetric 270° counterpart, as well as simulated maps of the intensity $I_{R(L)}(x,y)$ at points $(x,y)$ of the right- (left-) handed circularly polarized cathodoluminescence signal, spectrally integrated over the B and AB mode spectral windows and over the mirror collection angle.  Strong local CDS is clearly visible in both the experimental and simulated normalized dichroic signal $\gscav{\Omega_{exp}}(x,y)$ maps. Minor differences in resonant wavelength between experiments and simulations are attributed to the presence of a substrate in the experiments and the absence of correction for the spectrometer's wavelength-dependent quantum efficiency.
 
 Qualitatively, the experimental maps show weak dependence on mode energy and exhibit strong left- or right-circular polarization depending on which tip the electron beam is located on. The polarization sign reverses for the mirror-symmetric BKS (270$^{\circ}$). The signal is also polarized when the electron beam is positioned at the gap. However, no clear experimental trend emerged from inspection of different BKS. Simulation analysis reveals that the CDS signal at both tips of each individual nano-antenna has the same sign when the two antennas do not overlap spatially — a finding confirmed for the 270° BKS (see Figure \ref{fig:chiral_map}), where the experimental nano-antennas are more widely separated than in the corresponding simulations. The gap signal therefore depends on the precise spatial overlap of the two nano-antennas and will not be considered further when measuring $\Hel$. Focusing on the two external tips, the filtered maps for the 90° and 270° BKS are approximately mirror-symmetric, but are not chiral as they can be made to overlap by an in-plane rotation. This provides clear experimental confirmation that intensity alone cannot be reliably related to the helicity of the plasmonic modes.

 Beyond intensity differences, we studied the wavelength dependence of the CDS signal at both tips. The experimental results corresponding to the BKSs of Figure \ref{fig:chiral_map} clearly show that the maximum of the absolute value of the CDS signal is shifted in wavelength from one tip to the other for a given handedness (see Figure \ref{fig:setup}d), and has the same wavelength but opposite sign for a given tip peak for two opposite handednesses. This contrasts with the case of 2D chiral structures exhibiting a unique maximum, as corroborated by simulations and already demonstrated experimentally in \cite{Han2018RevealSubnanoscale}. This clearly points to the fact that the effect of chirality is indeed buried in the spatio-spectral symmetries of the BKS, that we will now explore.

\begin{figure}
    \centering
    \includegraphics[height=18cm]{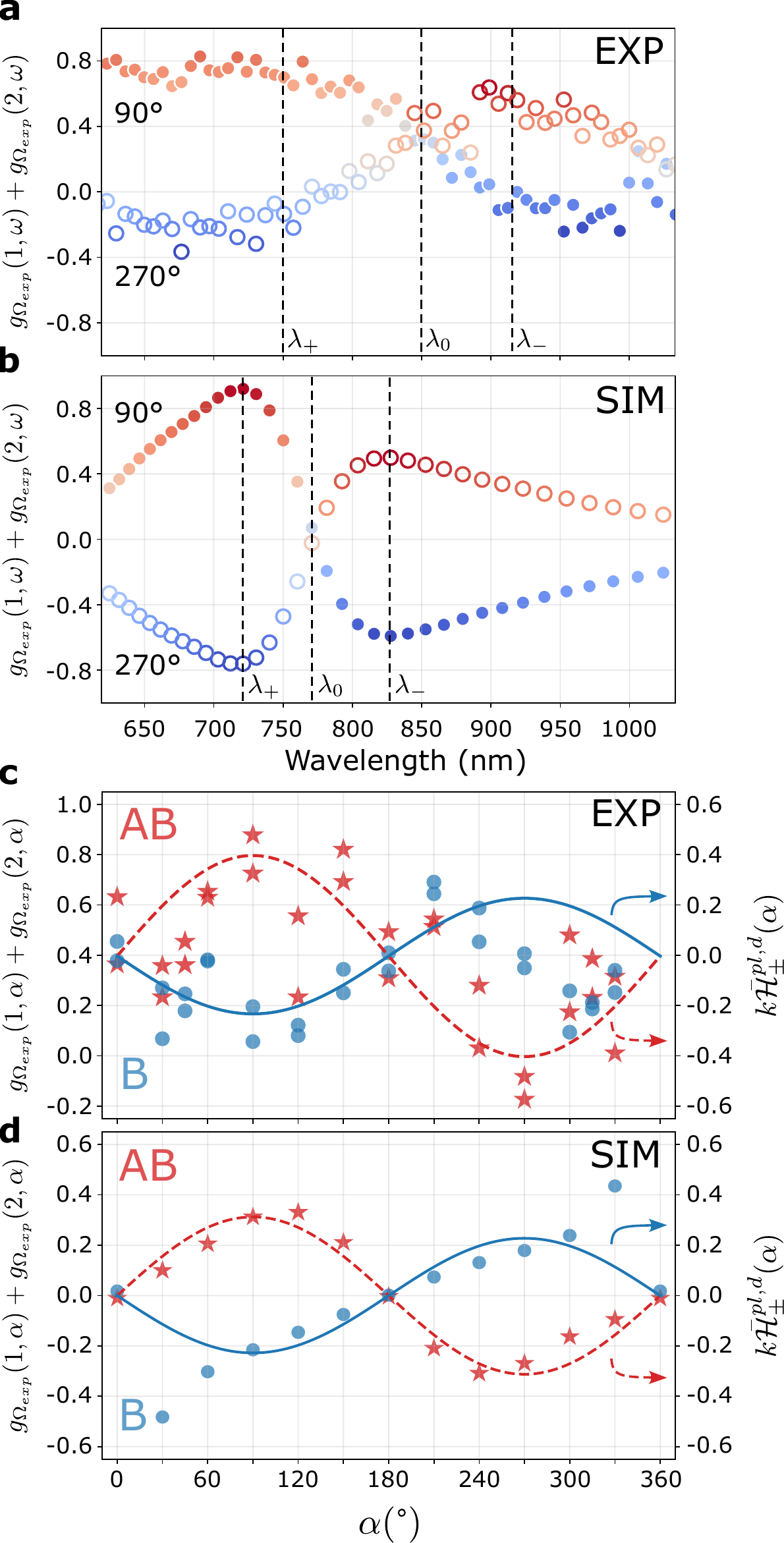}
    \caption{\textbf{Helicity measurement.} \textbf{(a)} Experimental  and \textbf{(b)} simulated (pyGDM) sum of the dichroic CL spectra acquired at the two tips  of the 90$^\circ$ and 270$^\circ$ BKS structures of Figure \ref{fig:chiral_map}. \textbf{(c)} Experimental and \textbf{(d)} simulated variations of the dichroic signal as a function of the BKS angle $\alpha$, filtered at the wavelengths of the bonding and antibonding mode summed on the two tips of the BKS. $k \Heldnorm$ as defined for the coupled dipole model is overlaid on the simulated data (\texttt{pyGDM}).}
    \label{fig:handedness}
\end{figure}

We present in Figure \ref{fig:handedness}a and b the measurements of the sum of the dichroic signal acquired on tip 1 and 2, $\gscav{\Omega_{exp}}(1)+\gscav{\Omega_{exp}}(2)$ and simulations of $\gscav{4\pi}$ for the 90$^\circ$ and 270$^\circ$ BKS. The simulations clearly reproduce the key features of a helicity, consistent with the two-dipole model. It changes sign at $\lambda_0$, and has opposite-sign optima at $\lambda_\pm$ that also change sign upon swapping handedness. Up to a constant offset, the experiments exhibit exactly the same behavior, considering that $\lambda_0$ and $\lambda_\pm$ have been determined experimentally and are therefore different from the simulated ones. We interpret this offset as the signature of the local environment asymmetry, the vertical antenna being supported by an $Si_3N_4$ substrate and embedded in $SiO_x$, on top of which the second antenna lies in vacuum. This clearly validates the fact that the two-tip sum is indeed proportional to the helicity in the case of the 90$^\circ$ and 270$^\circ$ BKS. In order to extend this analysis to other geometries (different angles $\alpha$), we define $\Hel_\pm(\alpha)$ ($\gscav{\Omega_{exp}}(\alpha)$) as the value of the $\Hel(\alpha,\omega)$ ($\gscav{\Omega_{exp}}(\alpha,\omega)$) integrated on a short window centered around the resonant energy of the bonding and antibonding modes, and divided by the integration width. Figure \ref{fig:handedness}c and d show experimental and simulated $\gscav{\Omega_{exp}}(\alpha)$, summed on the two tips, with the corresponding  $\Hel_{d,\pm}(\alpha)$ of the two-dipole model overlaid. As expected, $\Hel_{d,\pm}(\alpha)$ exhibits sinusoidal behaviour as a function of $\alpha$. The simulated sum over the two tips of $\gscav{\Omega_{exp}}(\alpha)$ also exhibits oscillatory behavior, which however departs for the bonding mode from the two-dipole scenario at angles $\alpha$ close to zero. This is expected because in this case the two antennas cannot be reduced to point dipoles anymore.   Here, we should recall that the two-dipole model is only an approximation of the two nano-antennas which requires refinement for realistic systems. Finally, the measured experimental sum on two tips of $\gscav{\Omega_{exp}}(\alpha)$ reproduces all the expected features of $\Hel_\pm(\alpha)$. This demonstrates that $\Hel$ can indeed be measured experimentally.

\section{Discussion}
As demonstrated, the principle for defining and measuring $\Hel$  is only based on general and symmetry-related considerations. While pCL is particularly well-suited for this measurement, the extension to other far-field or near-field techniques could in principle be envisaged, provided the excitation-sample-detection geometry can be symmetrized accordingly. This is straightforward for other linear techniques, where the equations for $g$ are exactly the same \cite{Guido2025}. In practice, however, implementation may prove challenging for certain techniques. The specific case of electron-based spectroscopy is instructive in this respect. For example, although photon-induced near field electron microscopy (PINEM) has demonstrated the capability to detect dichroism with high signal-to-noise ratio \cite{R.Harvey2020,Bourgeois2022}, and despite the formal analogy between PINEM and CL \cite{Auad2023}, the symmetrization required to retrieve the helicity may be difficult to achieve, as it would require a very large number of incident beam configurations. The methodology also unifies the  interpretation of the presence of the so-called "hidden chirality" \cite{Zu2018,Xie2025}, i.e, the appearance of a strong circular CDS signal whose sign changes seemingly arbitrarily across different parts of a sample. Previously invoked to explain this surprising, strong, and localized dichroic signal on achiral structures \cite{Hashiyada2014,Zu2018, Han2018RevealSubnanoscale, Horrer2020,Sannomiya2021,Mildner2023}, the same effect is here shown to dominate even the pCL for chiral objects. This explains why up to now pCL could not be directly used to measure chirality on chiral nanostructures \cite{Fang2016,Lingstadt2023}, unless we use a symmetry-based strategy as presented here to eliminate this dominant contribution.  

We have also derived analytical formulas for $\Hel$ in the case of a simplified two-dipole model ($\Held$). Based on the generality of our approximation and the close similarity of our simplified dipole model (equation \ref{EQ:Helicity_BKS}) to recent models using coupled dipoles to discuss chirality \cite{moser_conservation_2025,ossikovski_bornkuhn_2025}, we believe our methodology could be adapted to arbitrarily complex chiral objects (see SI S1B5). Moreover, although we have specialized this analysis to a plasmonic system, the fact that the helicity should be measurable through a properly symmetrized measurement of $g$ relies only on general considerations about the dual symmetry of the system, and not on which part of it (matter, electromagnetic field, or both) it applies. Therefore, our approach should be directly transferable to any polaritonic system.

As the hypotheses underlying the present definition of helicity and its measurement are particularly parsimonious (respect of the dual-symmetry in the measurement and existence of the helicity operator), we expect the extension to non-linear techniques to be valid as well.

Finally, our definition is of direct relevance to quantum plasmonics. The helicity is a fundamental scalar quantity in field theory, corresponding to the projection of a particle's spin onto its linear momentum. In the case of the photon, it corresponds to circular polarization $\pm 1$. When treating surface plasmons as polaritonic quasi-particles, our framework provides an unambiguous characterization of their helicity and, in particular, shows that, unlike photons, plasmons can take an infinite number of helicity values.

\section{Data Availability Statement}
The data that support the findings of this study are available from the corresponding author upon reasonable request.

\section{Acknowledgments}
This work is partly funded by the European Union’s Horizon 2020 research and innovation programme under grant  101017720 (EBEAM), the European Research Council advanced grant under grant 101201156 and by the Bernardo project of PEPR LUMA and was supported by the French National Research agency, as part of the France 2030 program, under grant ANR-24-EXLU-0001. The authors acknowledge financial support from the CNRS-CEA “METSA” French network (FR CNRS 3507) on the platform STEM-LPS CHROMATEM.

\section{Competing Interests}
MK is licensing the Mönch system used in this work to Attolight. The other authors declare no competing interests.

\section{Methods}
\subsection{Samples and experimental set-up}
 
For pCL experiments, BKS samples specially designed for electron microscopy were fabricated using a two-step electron-beam lithography process on a 15 nm thick silicon nitride membrane TEM grid (Ted Pella, Inc.). A standard electron-beam lithography and lift-off process was employed to create the bottom gold nano-antenna, each measuring approximately 150 nm in length, 50 nm in width, and 30 nm in thickness. A 10 nm thick silica spacer was then deposited by thermal evaporation at a 45° angle with a rotating sample holder, ensuring conformal coverage of all facets of the gold antenna. The final step consisted of a second electron-beam lithography process, in which alignment marks were used to precisely position the top gold nano-antenna over the bottom one, with the same dimensions.

Individual BKS structures can then be easily imaged by high-angle annular dark-field (HAADF) (see example of a 90$^\circ$ BKS in Figure \ref{fig:setup}a). As shown in the SI (S5), monochromated electron energy-loss spectroscopy (EELS)  clearly identified the spectral features of the two main B and AB modes (1.2 eV and 1.5 eV respectively) that are well separated due to strong coupling (Rabi energy on the order of 350 meV) \cite{Kociak2025} and their spatial variations, validating the quality and fitness to EM studies of the sample preparation. Higher order modes are discussed in the SI. A NION Hermes STEM microscope (CHROMATEM) working at 60 kV and 100 kV, fitted with a parabolic CL mirror from Attolight (Mönch) and a homemade CL analyzer (see SI section S6), as depicted in Fig.\ref{fig:setup}b, was used to simultaneously acquire at each electron beam position two CL spectra of opposite circular polarizations \cite{Baguenard2023}. Scanning the beam and recording the two spectra for each position resulted in two spatially resolved data cubes. The advantage of this parallel acquisition approach is to obtain a high-fidelity CDS signal with nanometer resolution without any drift-related noise in the case of sequential measurement of spectrum-images with different polarizations. The mirror collected about half the bottom hemisphere emission. All data were processed using HyperSpy \cite{hyperspy}.

\subsection{Simulations}
Simulations were performed using pyGDM2 \cite{wiecha2018pygdm,wiecha2022pygdm} and MNPBEM \cite{Hohenester2014}. All simulations were performed in the absence of a substrate, and for a constant overlap of the two antennas. This differs slightly from the experimental data where the overlap may vary from one BKS structure to another.

\PRLsep

\putbib[pCL]

\end{bibunit}

\fi

\if\SI1


\clearpage
\begin{bibunit}

\setcounter{page}{1}
\setcounter{section}{0}
\setcounter{figure}{0}
\setcounter{table}{0}
\setcounter{equation}{0}

\setcounter{secnumdepth}{3}
\renewcommand{\thesection}{S\arabic{section}}
\renewcommand{\thefigure}{S\arabic{figure}}
\renewcommand{\thetable}{S\arabic{table}}
\renewcommand{\theequation}{S\arabic{equation}}

\title{\sffamily\Large Supplemental material for: Absolute Experimental Measurement of Intrinsic Helicity in Elementary Plasmonic Nanostructures}
\date{\today}

\author{Malo Bézard}
\affiliation{%
Universit\'e Paris-Saclay, CNRS, Laboratoire de Physique des Solides, 91405, Orsay, France
}%

\author{Simon Garrigou}
 \affiliation{Univ Toulouse, CNRS, CEMES, Toulouse, France}%

 \author{Jérémie Béal}
\affiliation{%
Lumière, nanomatériaux, nanotechnologies (L2n), UMR CNRS 7076, Université de Technologie de Troyes, Troyes 10004, France
}%

 \author{Andreas Horrer}
\affiliation{%
Lumière, nanomatériaux, nanotechnologies (L2n), UMR CNRS 7076, Université de Technologie de Troyes, Troyes 10004, France
}%

\author{Yves Auad}
\affiliation{%
 Universit\'e Paris-Saclay, CNRS, Laboratoire de Physique des Solides, 91405, Orsay, France
}%

\author{Hugo Lourenço-Martins}
\email{hugo.lourenco-martins@cnrs.fr}
 \affiliation{Univ Toulouse, CNRS, CEMES, Toulouse, France}%

\author{Davy Gérard}
 \email{davy.gerard@utt.fr}
\affiliation{%
Lumière, nanomatériaux, nanotechnologies (L2n), UMR CNRS 7076, Université de Technologie de Troyes, Troyes 10004, France
}%

\author{Mathieu Kociak}
 \email{mathieu.kociak@universite-paris-saclay.fr}
\affiliation{%
 Universit\'e Paris-Saclay, CNRS, Laboratoire de Physique des Solides, 91405, Orsay, France
}%

\date{\today}

\maketitle

\clearpage
\onecolumngrid

\newcommand{\malo}[1]{\textcolor{blue}{#1}}

\maketitle

\tableofcontents
\newpage

\section{Plasmonic helicity and its application to the Born-Kuhn System (BKS)}
\label{sec:hel}

\subsection{General definition}

\noindent Optical chirality density is generally defined as \cite{tang_optical_2010-1}:
\begin{equation}\label{EQ:optical_chirality_definition}
    \mathcal{C}(\vec{r},t)= \frac{1}{2}\Big(\vec{E}.\vec{\nabla}\times \vec{E} + \vec{B}.\vec{\nabla}\times \vec{B}\Big) 
\end{equation}

\noindent where we have set the permittivity and permeability to $\mu_0=\epsilon_0=1$ for brevity. In this definition, one recognize the \textbf{helicity operator} $\hat{\mathcal{H}}$, which can be applied to any vectorial field $\vec{u}$ and read:
\begin{equation} \label{EQ:def_helicity}
    \hat{\mathcal{H}}[\vec{u}] = \vec{u}.\vec{\nabla}\times\vec{u}
\end{equation}

\noindent This operator is central in physics and apply in most of research fields. For example, if $\vec{u}$ represents a fluid velocity field, the previous equation corresponds to the so-called hydrodynamical helicity density \cite{wikipedia_nanophotonics}. Moreover, it is clear from \eqref{EQ:def_helicity} that the helicity of a field quantifies the ability of this field to form \emph{helices}. The optical chirality density as defined in \eqref{EQ:optical_chirality_definition}, is therefore simply the helicity density of the electric plus the helicity density of the magnetic field:
\begin{equation}\label{EQ:optical_chirality_definition}
    \mathcal{C}(\vec{r},t)= \frac{1}{2}\Big(\hat{\mathcal{H}}[\vec{E}] +\hat{\mathcal{H}}[\vec{B}] \Big) 
\end{equation}

\noindent and therefore measures the amount of helicoidal nature of the EM field lines. The \emph{total} helicity is then jus the integral in the whole space of this helicity, for any field $\vec{u}$:
\begin{equation}
    \mathcal{H} = \int d^3\vec{r} \; \hat{\mathcal{H}}[\vec{u}] = \int d^3\vec{r} \; (\vec{u}.\vec{\nabla}\times\vec{u})
\end{equation}

\noindent This way, one can define the \emph{total} optical chirality as:
\begin{equation} \label{EQ:def_chirality_tot}
    \mathcal{C}_\text{tot}(t) =  \int d^3\vec{r} \; \mathcal{C}(\vec{r},t)
\end{equation}

\noindent Importantly, the previous equation applied to a circularly polarized EM planewave will give $+1$ for a right-handed and $-1$ for a left-handed wave (up to a prefactor). \textbf{This shows that, loosely speaking, the optical chirality can be seen as a generalized local polarization for non planewave-like fields}.\\

\noindent With those definitions in hands, our goal in this section is to find an operational definition of the "chirality" of a localized surface plasmon (LSP). A "naive" approach would be to compute \eqref{EQ:def_chirality_tot} for the eigenfields $\vec{E}_n,\vec{B}_n$ of the n$^\text{th}$ plasmon mode. However, this approach:
\begin{enumerate}
    \item Is computationally heavy and does not necessarily allow closed analytical formula. 
    \item does not tackle a central and topical problem which is the proper definition of the optical chirality in presence of a dispersive or dissipative medium. In particular, a polariton (like a LSP) is field resulting from the coupling between the EM field and a matter field (here charge density waves) and it is expected that the total helicity of the polaritonic field must be - at least partly - stored in the matter \cite{alpeggiani_electromagnetic_2018, vazquez-lozano_optical_2018, fernandez-corbaton_total_2021, mackinnon_quantized_2025}.
\end{enumerate}

\noindent Since LSP are essentially matter waves (contrary to surface plasmon polariton which behave as confined light waves), we postulate that their chirality must be described by matter degrees of freedom only. It is clear from the above-mentioned definitions that we have an obvious choice for our defining  the chirality of plasmons: employing the helicity operator. In particular, guided by the Hopfield model, a LSP is fully characterized by its density of polarization $\vec{\mathfrak{p}}(\vec{r},t)$ and therefore, we define: 
\begin{equation} \label{EQ:helicity_plasmon_continu}
\boxed{
    \mathcal{H}^\text{pl}(t) = \int \vec{\mathfrak{p}}(\vec{r},t).\vec{\nabla}.\vec{\mathfrak{p}}(\vec{r},t) d^3\vec{r} = \int \hat{\mathcal{H}}[\vec{\mathfrak{p}}(\vec{r},t)]d^3\vec{r}
    }
\end{equation}

\noindent which is the total helicity of the plasmonic polarization field $\vec{\mathfrak{p}}(\vec{r},t)$. As for the optical and hydrodynamical cases discussed above, $\mathcal{H}^\text{pl}$ the helicoidal nature of the polarization field. \textbf{As it is detailed in the main text, the term \emph{chiraliy} is polysemous, and to remove any ambiguity, we will call $\mathcal{H}^\text{pl}$ the (total) helicity of the plasmon field}. It also stresses the fact that only the helicity operator is involved in our definition. The term chirality will only be used to characterize the structure of the nano-particle underlying the plasmon.\\

\noindent In the next sections, \textbf{we will show why this definition makes sense and how it can be related to experimental observables.}

\subsection{Two static dipoles} \label{SSEC:static_dipoles_chirality}

\noindent To make sense of our previous definition, we will turn to a simple but important case: the Born-Kuhn system (BKS). It is made of two nano-particles, each underlying a dipolar LSP resonance represented by two dipoles $\vec{\mathfrak{p}}_1,\vec{\mathfrak{p}}_2$ located at $\vec{r}_1,\vec{r}_2$. \textbf{To start, we suppose that the two dipole are static so that the time-dependence in $\mathfrak{p}(\vec{r},t)$ disappears (the time-dependent case will be treated in the next section)}. Intuitively, it is the most extreme case of the formula \eqref{EQ:helicity_plasmon_continu} where the polarization density can be reduced to (This definition can straightforwardly be extended to an arbitrary number of dipoles.):
\begin{equation}
  \vec{\mathfrak{p}}(\vec{r})=\underbrace{\vec{\mathfrak{p}}_1 \; f_\varepsilon (\vec{r}-\vec{r}_1)}_{=\mathfrak{p}_1(\vec{r})}+\underbrace{\vec{\mathfrak{p}}_2 \; f_\varepsilon(\vec{r}-\vec{r}_2)}_{=\mathfrak{p}_2(\vec{r})}
  \label{EQ:smoothed_plasmon_field}
\end{equation}

\noindent Here, $f_\varepsilon$ is a sequence of smooth functions satisfying $\lim_{\varepsilon\rightarrow 0} f_\varepsilon =\delta(x)$; therefore it also verifies the (necessary) symmetry property $f_\varepsilon(\vec{r})=f_\varepsilon(-\vec{r})$. For the time being, we forget the time-dependence of the two dipoles. We will include it properly in section \ref{SEC:modele_couple} where the coupling (hence the phase-relation) between the two dipoles will be properly derived. We note that definition of a plasmon is somehow arbitrary: it can relate to the charges or dipoles \textit{within} the particle(s) subtending it, or the electromagnetic field \textit{induced} by these charges or dipoles outside of the particle(s). As we will see, the use of dipoles as a definition for plasmons is an efficient way to characterize the helicity. It has then to be bear in mind than the plasmonic field in equation \eqref{EQ:smoothed_plasmon_field} is not the induced electromagnetic field, but the dipolar field inside the nanoantenna (point-dipoles here).\\

\noindent With this minimal description of the BKS plasmon \eqref{EQ:smoothed_plasmon_field}, we are now in position to compute it helicity \eqref{EQ:helicity_plasmon_continu}. To do so, we start with a algebraic representation of the \emph{handedness pseudotensor density} (HPD) $\chi$ from \cite{efrati_orientation-dependent_2014}:
\begin{equation}\label{EQ:HPD_tensorial}
  n^i \chi_{ij} m^j = n^i \partial_i P^k \epsilon_{jlk} P^l m^j
\end{equation}

\noindent where Einstein's convention for implicit summation over repeated indices is assumed and $\epsilon_{jlk}$ is the Levi-Civita tensor, $\vec{n}$ is an orientation vector, $\vec{m}$ a rotation pseudovector and $P^l$ the $l^\text{th}$ component of $\vec{P}(\vec{r},t)$. From this HPD, one can then derive the \emph{full handedness pseudotensor}: 
\begin{equation}
  \mathcal{X}_{ij}= \int  \chi_{ij} d^3\vec{r} 
\end{equation}

\noindent And finally, the trace of the previous equation then corresponds to the previously defined plasmonic helicity:
\begin{equation}
    \mathcal{H}^\text{pl}=\Tr[\mathcal{X}]
\end{equation}

\noindent We thus start by plugging \eqref{EQ:smoothed_plasmon_field} in \eqref{EQ:HPD_tensorial}, one obtains:
\begin{equation}
    \chi_{ij}= \partial_i[\vec{\mathfrak{p}}^k_1 \; f_\varepsilon (\vec{r}-\vec{r}_1)+\vec{\mathfrak{p}}^k_2 \; f_\varepsilon(\vec{r}-\vec{r}_2)]\epsilon_{jlk} [\vec{\mathfrak{p}}^l_1 \; f_\varepsilon (\vec{r}-\vec{r}_1)+\vec{\mathfrak{p}}^l_2 \; f_\varepsilon(\vec{r}-\vec{r}_2)]
\end{equation}

\noindent A brute force expansion gives 4 terms:
\begin{equation}
  \begin{split}
    \chi_{ij}=&\textcolor{red}{\partial_i\left\{\vec{\mathfrak{p}}^k_1 f_\varepsilon (\vec{r}-\vec{r}_1) \right\} \epsilon_{jlk} \, \vec{\mathfrak{p}}^l_1 \; f_\varepsilon (\vec{r}-\vec{r}_1) + \partial_i\left\{\vec{\mathfrak{p}}^k_2 f_\varepsilon (\vec{r}-\vec{r}_2) \right\} \epsilon_{jlk} \, \vec{\mathfrak{p}}^l_2 \; f_\varepsilon (\vec{r}-\vec{r}_2)}\\
    &+ \textcolor{blue}{\partial_i\left\{\vec{\mathfrak{p}}^k_1 f_\varepsilon (\vec{r}-\vec{r}_1) \right\} \epsilon_{jlk} \, \vec{\mathfrak{p}}^l_2 \; f_\varepsilon (\vec{r}-\vec{r}_2)+ \partial_i\left\{\vec{\mathfrak{p}}^k_2 f_\varepsilon (\vec{r}-\vec{r}_2) \right\} \epsilon_{jlk} \, \vec{\mathfrak{p}}^l_1 \; f_\varepsilon (\vec{r}-\vec{r}_1)}
    \end{split}
\end{equation}

\noindent where, for clarity, we highlighted in \textcolor{red}{red the single dipole terms} and in \textcolor{blue}{blue the crossed terms}. The total handedness pseudotensor then reads:
\begin{equation}
   \mathcal{X}_{ij}= \epsilon_{jlk} \Big( \vec{\mathfrak{p}}^k_1 \, \vec{\mathfrak{p}}^l_1 \; I^{11} _i + \vec{\mathfrak{p}}^k_2 \, \vec{\mathfrak{p}}^l_2 \; I^{22} _i +  \vec{\mathfrak{p}}^k_2 \, \vec{\mathfrak{p}}^l_1 \; I^{21} _i +  \vec{\mathfrak{p}}^k_1 \, \vec{\mathfrak{p}}^l_2 \; I^{12}_i \Big)
\end{equation}

\noindent where the $I_j$ represents integrals of the form:
\begin{equation}
  I^{12}_i = \int d^3\vec{r} f_\varepsilon (\vec{r}-\vec{r}_2) \partial_i f_\varepsilon (\vec{r}-\vec{r}_1) 
\end{equation}

\subsubsection{Single dipole terms}

\noindent Let's define the tensor:
\begin{equation}
  P_{AB}^{kl}=\mathfrak{p}_A^k \mathfrak{p}_B^l
\end{equation}

\noindent where $A,B=1,2$ indexes the dipoles. Then:
\begin{equation}
   \mathcal{X}_{ij}= \epsilon_{jlk} \Big( P_{11}^{kl} \; I^{11} _i + P_{22}^{kl} \; I^{22} _i + P_{21}^{kl}  \; I^{21} _i +  P_{12}^{kl}  \; I^{12}_i \Big)
\end{equation}

\noindent Besides, it is clear from its definition that the $P$ tensor is symmetric if $A=B$, for example:
\begin{eqnarray}
  P_{11}^{lm}&=&P_{11}^{ml}\\
  P_{11}^{lm}-P_{11}^{ml} &=& 0\\
  (\delta_{il}\delta_{jm}-\delta_{im}\delta_{jl})P_{11}^{ij}&=&0\\
  \epsilon_{klm}\epsilon_{ijk}P_{11}^{ij}&=&0
\end{eqnarray}

\noindent which implies that:
\begin{equation}
  \epsilon_{jkl}P_{AA}^{kl}=0
\end{equation}

\noindent The \textbf{single dipole terms are thus null}. The handedness pseudotensor can therefore be re-written  as a function of the crossed terms only:
\begin{equation}
   \mathcal{X}_{ij}= \epsilon_{jlk} \Big( P_{12}^{kl}  \; I^{12}_i + P_{21}^{kl}  \; I^{21} _i \Big)
\end{equation}

\subsubsection{Simplification of the crossed terms}

\noindent The anti-symmetry of the Levi-Civita symbol leads to:
\begin{equation}
  \epsilon_{jkl}P^{kl}_{12}=-\epsilon_{jkl}P^{kl}_{21}
\end{equation}

\noindent Therefore:
\begin{equation}
  \mathcal{X}_{ij}=  \epsilon_{jkl}P^{kl}_{12} \Big(I^{12}_i-I^{21}_i\Big)
\end{equation}

\noindent Now we examine the term:
\begin{equation}
   I^{12}_i = \int d^3\vec{r} f_\varepsilon (\vec{r}-\vec{r}_2) \partial_i f_\varepsilon (\vec{r}-\vec{r}_1) 
\end{equation}

\noindent we now define $\vec{r}_{12}=\vec{r}_{2}-\vec{r}_{1}$. Changing the integration variable to $\vec{x}=\vec{r}-\vec{r}_2$ leads to:
\begin{equation}
  I^{12}_i = \int d^3\vec{x} f_\varepsilon (\vec{x}) \partial_i f_\varepsilon (\vec{x}+\vec{r}_{12}) \equiv  I^{+}_i
\end{equation}

\noindent Similarly, using $\vec{x} = \vec{r}-\vec{r}_1$:  
\begin{equation}
  I^{21}_i = \int d^3\vec{x} f_\varepsilon (\vec{x}) \partial_i f_\varepsilon (\vec{x}-\vec{r}_{12}) \equiv  I^{-}_i
\end{equation}

\noindent Besides, an integration by part of the previous equation gives:
\begin{eqnarray}
   I^{+}_i &=& \Big[ f_\varepsilon (\vec{x}) f_\varepsilon (\vec{x}+\vec{r}_{12})  \Big]_{\vert \vec{x}\vert\rightarrow\infty}-\int d^3\vec{x} f_\varepsilon (\vec{x}+\vec{r}_{12}) \partial_i f_\varepsilon (\vec{x})\\
   &=&-I^{-}_i
\end{eqnarray}

\noindent Therefore:
\begin{equation}
  \mathcal{X}_{ij}= 2\epsilon_{jkl}\; P^{kl}_{12} \; I^{+}_i
\end{equation}

\noindent Let's now define the function:
\begin{equation}
  F(\vec{r}_{12})= \int d^3 \vec{x} f_{\epsilon}(\vec{x}) f_{\epsilon}(\vec{x}-\vec{r}_{12})
  \label{eq:F}
\end{equation}

\noindent Then, taking the partial derivative with respect to the variable $\vec{r_{12}}$:
\begin{eqnarray}
   \partial^{r_{12}}_{i}F(\vec{r}_{12})&=& \int d^3 \vec{x} f_{\epsilon}(\vec{x}) \partial^{r_{12}}_{i} f_{\epsilon}(\vec{x}-\vec{r}_{12})
\end{eqnarray}

\noindent Besides:
\begin{equation}
   \partial^{r_{12}}_{i} f_{\epsilon}(\vec{x}-\vec{r}_{12}) = - \partial^{x}_{i} f_{\epsilon}(\vec{x}-\vec{r}_{12})
\end{equation}

\noindent and therefore:
\begin{eqnarray}
   \partial^{r_{12}}_{i}F(\vec{r}_{12})&=& - \int d^3 \vec{x} f_{\epsilon}(\vec{x}) \partial^{x}_{i} f_{\epsilon}(\vec{x}-\vec{r}_{12}) = -I^{-}_i = I^{+}_i 
\end{eqnarray}

\noindent Hence:
\begin{equation}
  \mathcal{X}_{ij}= 2\epsilon_{jkl}\; P^{kl}_{12} \;  \partial^{r_{12}}_{i}F(\vec{r}_{12})
\end{equation}

\noindent Besides, we now suppose that $f_\epsilon(\vec{r})$ is isotropic, i.e. only depends on $\Vert \vec{r} \Vert $ which is compatible with its definition. In that case, $F$ is also isotropic and therefore:
\begin{equation}
  \partial^{r_{12}}_{i}F(\vec{r}_{12}) = F'(\vec{r}_{12}) \; \hat{r}_{i,12}
\end{equation}

\noindent with $\hat{r_{i,12}}=\vec{r}_{i,12}/\Vert \vec{r}_{i,12} \Vert$. Which eventually gives:
\begin{equation}
  \mathcal{X}_{ij}= 2\frac{F'(\vec{r}_{12})}{\Vert \vec{r}_{12}\Vert}\,\epsilon_{jkl}\; \mathfrak{p}_1^k \mathfrak{p}_2^l \; (\vec{r}_{2,i}-\vec{r}_{1,i})
\end{equation}

\subsubsection{Helicity of the plasmon mode}

\noindent Now, since $\epsilon_{jkl}\; \mathfrak{p}_1^k \mathfrak{p}_2^l=(\mathfrak{p}_1 \times \mathfrak{p}_2)_j$, taking the trace of the previous tensor gives (note the  sign change due to the  inverted 1 and 2 in the $\vec{r}$ bracket):
\begin{equation}
  \mathcal{H}^\text{pl}=\Tr [\mathcal{X}]= -2\frac{F'(\vec{r}_{12})}{\Vert \vec{r}_{12}\Vert}\, (\vec{\mathfrak{p}}_1\times \vec{\mathfrak{p}}_2).(\vec{r}_1-\vec{r}_2)
\end{equation}

\noindent One can see that this continuous pseudo-scalar obviously goes to zero if $(\vec{r}_1-\vec{r}_2)\rightarrow 0$ or the angle $\alpha$ between both dipoles tends to 0. Moreover, it changes sign when exchanging the dipoles positions or dipoles configuration (bounding or anti-bonding modes). All these properties are reassuring indications that this is the relevant helicity for LSP. 

\subsubsection{Discussion on regularization method}

\noindent We note that the definition of the helicity given above suffers from a regularization issue. Indeed, we have  demonstrated using the universal definition of \cite{efrati_orientation-dependent_2014} that  $ \mathcal{H}^\text{pl}$ is a measurement of the helicity of the plasmon modes, as:
\begin{equation}
  \mathcal{H}^\text{pl}=\Tr [\mathcal{X}] = \aleph(\epsilon) \; (\vec{\mathfrak{p}}_1\times \vec{\mathfrak{p}}_2).(\vec{r}_1-\vec{r}_2)
\end{equation}

\noindent with:
\begin{equation}\label{EQ:deg_aleph}
  \aleph(\epsilon) = -2\frac{F'(\vec{r}_{12})}{\Vert \vec{r}_{12}\Vert}
\end{equation}

\noindent If we now choose a Gaussian sequence as our $f_\epsilon$:
\begin{equation}
  f_\epsilon (\vec{r})=\frac{e^{-\tfrac{r^2}{2\epsilon^2}}}{(2\pi\epsilon^2)^{3/2}}
\end{equation}

\noindent Computing the convolution product \eqref{eq:F} will essentially give $F\sim \exp(-r_{12}^2)/(4\epsilon^2)$ (2 times the width when convoluting two Gaussians). Then, $F'\sim (-r/2\epsilon^2)F$ and thus:
\begin{equation}
  \aleph(\epsilon) \sim -\frac{1}{2\epsilon^3} e^{\tfrac{-r_{12}^2}{4\epsilon^2}}
\end{equation}

\noindent for a finite $r_{12}$, we therefore have $\lim_{\epsilon\rightarrow 0} \aleph(\epsilon) = 0$. In other words, in the point limit of our field $\vec{u}$, this helicity measure vanishes. This is expected as we are applying a continuous model \cite{efrati_orientation-dependent_2014} to a discrete dipole distribution - which cannot be treated as a field \textit{stricto sensu}. This is not a limit of the definition of \cite{efrati_orientation-dependent_2014} but a limit of our model, which reduces plasmonic rod to point-like dipoles. Our regularization is close to what one could encounter in lattice models where a minimal length (here $\epsilon$) is introduced. Therefore, choosing a very small but finite $\epsilon$, we obtain a regularized measure of the helicity of the plasmon mode:
\begin{equation} \label{EQ:static_plasmon_helicity}
  \boxed{
  \mathcal{H}^\text{pl} = \aleph(\vec{r}_{12}) \, (\vec{\mathfrak{p}}_1\times \vec{\mathfrak{p}}_2).(\vec{r}_1-\vec{r}_2)
  }
\end{equation}

\subsubsection{Generalization to $n$ dipoles}

\noindent The previous equation can be generalized to:
\begin{equation}
    \mathcal{H}^\text{pl}=\sum_{n=1}^N \sum_{m\neq n}^N \aleph(r_{nm}) (\vec{\mathfrak{p}}_n \times \vec{\mathfrak{p}}_m).(\vec{r}_n-\vec{r}_m)
    \label{EQ:general_plasmon_helicity}
\end{equation}

\noindent Equation (\ref{EQ:static_plasmon_helicity}) is the fomula giving the helicity associated to the BKS in the temporal domain. Let us call it $\mathcal{H}^\text{pl, BKS}$. Using this notation we can rewrite (\ref{EQ:general_plasmon_helicity}) as :

\begin{equation}
    \mathcal{H}^\text{pl}=\sum_{n=1}^N \sum_{m\neq n}^N \mathcal{H}^\text{pl, BKS}_{nm} = 2\sum_{n=1}^N\sum_{m>n}^N\mathcal{H}^\text{pl, BKS}_{nm}
    \label{EQ:general_plasmon_helicity_BKS}
\end{equation}

\noindent This equation provides us a representation of the general behavior of a BKS, since, according to (\ref{EQ:general_plasmon_helicity_BKS}), the BKS can be used to calculate the helicity of any arbitrary structure, provided that the structure can be decomposed into a sum of dipoles.

\subsection{Oscillating dipoles}

\noindent Let's now imagine that our two dipoles $p_j, j=1,2$ are not static but oscillating harmonically at frequency $\omega$:
\begin{equation}
    \vec{\mathfrak{p}}_j(t)=\Re \{ \vec{p}_j(\omega)e^{-i\omega t} \} = \frac{1}{2} \left[ \vec{p}_j(\omega)e^{-i\omega t} + \vec{p}^*_j(\omega)e^{i\omega t}  \right]
\end{equation}

\noindent where $\vec{p}_j(\omega)$ denotes the Fourier component of $\vec{\mathfrak{p}}_j(t)$ at frequency $\omega$. In order to generalize the definition \eqref{EQ:static_plasmon_helicity} to oscillating dipoles, we need to compute the time-averaged $\braket{\;}_T$ value of the helicity over one optical cycle of period $T$, i.e.:
\begin{equation}
    \braket{\mathcal{H}^\text{pl}}_T = \aleph \, \braket{\vec{\mathfrak{p}}_1(t)\times \vec{\mathfrak{p}}_2(t)}_T \;. \;(\vec{r}_1-\vec{r}_2)
\end{equation}

\noindent since the position of the dipoles is fixed. We thus need to compute:
\begin{align}
\braket{\vec{\mathfrak{p}}_1(t)\times \vec{\mathfrak{p}}_2(t)}_T
&= \frac{1}{T}\int_0^T \vec{\mathfrak{p}}_1(t)\times \vec{\mathfrak{p}}_2(t)\, dt \\
&= \frac{1}{4T}\int_0^T
\left[ \vec{p}_1(\omega)e^{-i\omega t}
+ \vec{p}^*_1(\omega)e^{i\omega t} \right]
\left[ \vec{p}_2(\omega)e^{-i\omega t}
+ \vec{p}^*_2(\omega)e^{i\omega t} \right] dt \\
&= \frac{1}{4T}\int_0^T \Big[
\vec{p}_1(\omega)\times \vec{p}_2(\omega) e^{-2i\omega t}
+ \vec{p}^*_2(\omega)\times \vec{p}^*_1(\omega) e^{2i\omega t} \\
&\quad\quad
+ \vec{p}_1(\omega)\times \vec{p}^*_2(\omega)
+ \vec{p}^*_1(\omega)\times \vec{p}_2(\omega)
\Big] dt
\end{align}

\noindent The two first terms will average at 0 and therefore:
\begin{eqnarray}
    \braket{\vec{\mathfrak{p}}_1(t)\times \vec{\mathfrak{p}}_2(t)}_T &=& \frac{1}{4} \left[ \vec{p}_1(\omega)\times \vec{p}^*_2(\omega) + \vec{p}^*_1(\omega)\times \vec{p}_2(\omega)\right]\\
    &=& \frac{1}{2} \Re \left\{\vec{p}_1(\omega)\times \vec{p}^*_2(\omega) \right\}
\end{eqnarray}

\noindent Therefore:
\begin{equation} \label{EQ:harmonic_plasmon_helicity}
\boxed{
    \braket{\mathcal{H}^\text{pl}}_T = \frac{\aleph}{2} \, \Re \left\{\vec{p}_1(\omega)\times \vec{p}^*_2(\omega) \right\}. (\vec{r}_1-\vec{r}_2)
    }
\end{equation}

\noindent \textbf{We will indifferently denote the quantities \eqref{EQ:static_plasmon_helicity} and \eqref{EQ:harmonic_plasmon_helicity} as $\mathcal{H}^\text{pl}$, keeping in mind that this time-averaging procedure is implicit when applied to an oscillating dipole}. For a more general continuous polarization density $\vec{p}(\vec{r},t)$, the previous equation generalizes as:
\begin{equation}
\boxed{
    \braket{\mathcal{H}^\text{pl}}_T = \frac{1}{2} \Re{ \Big\{ \int_{\mathbb{R}^3} [\vec{p}^*(\vec{r},\omega).\vec{\nabla}\times\vec{p}(\vec{r},\omega)] d^3\vec{r}} \Big\}
    }
\end{equation}

\subsection{Connection with the standard definition of dipole chirality}

\noindent In the literature, a chiral dipole is generally defined by $\Im\{\vec{p}.\vec{m}^*\}\neq 0$. Let's compute this quantity for the previously developed two dipoles model $\vec{p}_n, n\in\{1,2 \}$. The total electric dipole simply is:
\begin{equation}
    \vec{p}_\text{tot}=\vec{p}_1+\vec{p}_2
\end{equation}

\noindent To each oscillating dipole can be associated a current density:
\begin{equation}
    \vec{j}_n=-i\omega \vec{p}_n
\end{equation}

\noindent Besides the magnetic moment can be defined as:
\begin{equation}
    \vec{m}=\frac{1}{2}\int \vec{r}\times \vec{j}(\vec{r}) d^3\vec{r}
\end{equation}

\noindent which for the two dipoles reduces to:
\begin{equation}
    \vec{m}=\frac{1}{2}\sum_n \vec{r}_n\times \vec{j}_n = -\frac{i\omega}{2}(\vec{r}_1\times\vec{p}_1+\vec{r}_2\times\vec{p}_2)
\end{equation}

\noindent We can now compute the scalar product:
\begin{equation}
    \vec{p}.\vec{m}^* = \frac{i\omega}{2}(\vec{p}_1+\vec{p}_2).(\vec{r}_1\times\vec{p}^*_1+\vec{r}_2\times\vec{p}^*_2)
\end{equation}

\noindent Let's choose the coordinate system so that $\vec{r}_2=0$ and $\vec{R}=\vec{r}_1-\vec{r_2}$:
\begin{equation}
      \vec{p}.\vec{m}^* = \frac{i\omega}{2} \vec{p}_1.(\vec{R}\times \vec{p}_1^*)+\frac{i\omega}{2} \vec{p}_2.(\vec{R}\times \vec{p}_1^*)
\end{equation}

\noindent Taking the imaginary part:
\begin{equation}\label{EQ:magnetic_chirality_step}
    \Im\{\vec{p}.\vec{m}^*\} = \frac{\omega}{2} \Re\{\vec{p}_1.(\vec{R}\times \vec{p}_1^*)\}+\frac{\omega}{2} \Re\{\vec{p}_2.(\vec{R}\times \vec{p}_1^*)\}
\end{equation}

\noindent Besides, using the cyclic property of the scalar triple product $a.(b\times c)=b.(c\times a) = c.(a\times b)$ and the antisymmetry of the cross produc $a\times b=-b\times a$, we obtain:
\begin{equation}
   (\vec{p}_1.(\vec{R}\times \vec{p}_1^*))^*=\vec{p}^*_1.(\vec{R}\times \vec{p}_1) = - \vec{p}_1.(\vec{R}\times \vec{p}_1^*)
\end{equation}

\noindent In other words, $\vec{p}_1.(\vec{R}\times \vec{p}_1^*)\in i\mathbb{R}$, therefore the first term in \eqref{EQ:magnetic_chirality_step} vanishes. Applying again the cyclic property of the scalar triple product to the second term, we finally get:
\begin{equation}
      \Im\{\vec{p}.\vec{m}^*\} = \frac{\omega}{2} \Re\{\vec{R}.(\vec{p}_1^*\times\vec{p}_2)\} = \frac{\omega}{2} \Re\{\vec{R}.(\vec{p}_1\times\vec{p}_2^*)\}
\end{equation}

\noindent Comparing with \eqref{EQ:harmonic_plasmon_helicity}, we obtain:
\begin{equation}
\boxed{
    \braket{\mathcal{H}^\text{pl}}_T = \frac{\aleph}{\omega}\Im\{\vec{p}.\vec{m}^*\}
    }
\end{equation}

\noindent This calculation shows how a chiral dipole can emerge from two shifted and rotated electric dipoles which produce an effective dipole moment - even without an intrinsic magnetic polarizability - a well-known result usually called the two-group model, see \cite{andrews_quantum_2018} and the references therein.

\section{Coupled point dipoles model for polarized cathodoluminescence } \label{SEC:modele_couple}

\subsection{Far-field dichroism of two coupled dipoles excited by arbitrary electrical fields}

In order to model a BKS made up of two plasmonic antennas, we model the later as oscillating points dipoles $\vec{p}_1$ and $\vec{p}_2$ at positions $\vec{r}_1$ and $\vec{r}_2$ given by:
$\mathbf{r_1} = (0, a, -d)$ and $\mathbf{r_2} = (a\sin(\alpha), a\cos(\alpha), 0)$ and orientations $\vec{e}_1$ and $\vec{e}_2$ which reads in the (x,y,z) basis: 

\begin{equation}
    \begin{split}
        &\vec{e}_1 = \begin{pmatrix}
1\\
0 \\
0
\end{pmatrix}\\
        &\vec{e}_2 = \begin{pmatrix}
\cos(\alpha)\\
\sin(\alpha) \\
0
\end{pmatrix}
    \end{split}
\end{equation}

\noindent where $a$ is the dipole position along $\vec{e}_1$ and $\vec{e}_2$, $d$ the distance between the two dipoles along $z$ and $\alpha$ the angle between the two dipoles in the $(x,y)$ plane.  We note that we have explicitly considered pairs of dipoles parallel to the $(x,y)$ plane to stick to the BKS model. The model could easily be expanded to the case of two dipoles not being in parallel planes. Each point dipole posses the same resonant frequency $\omega_0$ and dissipation term $\gamma$, and both are coupled thanks to a coupling term $\omega_c$. We define:

\begin{equation} \label{EQ:def_Omega_squared}
\Omega^2 = \omega_0^2 - i\gamma\omega - \omega^2
\end{equation}

\begin{equation}
T^2 = \frac{\omega_p^2}{\Omega^4 - \omega_c^4} \in \mathbb{C}, \quad \omega_p^2 = \frac{e^2}{m}
\end{equation}

\noindent The goal of this section is to compute the electric energy radiated in the far-field by the BKS system. In the following, the electric far-field amplitude in a certain solid angle $S_\Omega$ will be noted $\vec{F}(S_\Omega)$ and its projection onto the circular polarization basis represented by the unit vectors $\vec{u}_\pm = (1/\sqrt{2})(\hat{x} \pm i \hat{y})$ will be noted $F_\pm (S_\Omega)=\vec{F}(S_\Omega).\vec{u}_\pm$. In spherical coordinates, we have $F_\pm=(1/\sqrt{2})(F_\theta\pm i F_\phi)$. With these notations, the $S_3$ Stokes component that quantifies the difference between right- and left-handed amplitudes in the far-field is given by:
\begin{equation}
S_3(S_\Omega) = \abs{F_+(S_\Omega)}^2 - \abs{F_{-}(S_\Omega)}^2
\end{equation}

\noindent This definition is naturally valid for any incident electrical field $\vec{E}$ driving the BKS system.\\

\noindent We now note the projection of this arbitrary exciting field $\vec{E}(\vec{r})$ onto the two dipoles $\vec{e}_{1,2}$ as $E_{1,2}=\vec{E}(\vec{r}).\vec{e}_{1,2}$. With this notations:
\begin{equation}\label{eq:ch4:BKS_pheno_dipole_expressions}
    \left\{\begin{split}
&\bvec{p}_1  = T^2 \Big[ \Omega^2 E_{1}(\vec{r}_1) - \omega_c^2 E_{2}(\vec{r}_2) \Big] \bvec{e}_1 & = p_1 \bvec{e}_1 \\
&\bvec{p}_2  = T^2 \Big[ -\omega_c^2 E_{1}(\vec{r}_1) + \Omega^2 E_{2}(\vec{r}_2) \Big] \bvec{e}_2 & = p_2 \bvec{e}_2
\end{split}\right.
\end{equation}

\noindent The far-field amplitude generated by the dipole system can be straightforwardly calculated by applying the asymptotic Green dyadic:
\begin{equation}
    \vec{F} = \tensorarrow{G}_\infty(\bvec{r}_1,\bvec{k},r_\infty,\omega) \cdot \bvec{p}_1 + \tensorarrow{G}_\infty(\bvec{r}_2,\bvec{k},r_\infty,\omega) \cdot \bvec{p}_2
\end{equation}

\noindent where $\vec{k}$ denotes the far-field wavevector of modulus $k=\omega/c$ pointing at an infinitesimal solid angle $S_\Omega$ and $r_\infty$ is the far-field sphere's radius. The asymptotic Green dyadic can be decomposed as:
\begin{equation}
    \tensorarrow{G}_\infty(\bvec{r}_1,\bvec{k},r_\infty,\omega) = A(k,r_\infty) e^ {i\phi(\bvec{r}_1, \vec{k})}\tensorarrow{\mathfrak{g}}(\hat{k})
\end{equation}

\noindent with the phase term:
 \begin{equation}
\phi(\bvec{r},\vec{k}) = \vec{k} \cdot \bvec{r} 
\end{equation}

\noindent the amplitude:
\begin{equation}
A(k,r_\infty) := \frac{k^2 e^{i k r_\infty}}{r_\infty}
\end{equation}

\noindent and the dyadic geometrical radiation term:
\begin{equation}
  \tensorarrow{\mathfrak{g}}(\hat{k})=  \begin{pmatrix}
        1-\sin^2(\theta) \cos^2 (\varphi)  & -\cos (\varphi) \sin (\varphi) \sin^2(\theta)  & -\cos(\theta)\sin(\theta)\cos(\varphi)\\
        -\cos (\varphi) \sin (\varphi) \sin^2(\theta) &  1-\sin^2(\theta) \sin^2 (\varphi) & -\cos(\theta)\sin(\theta)\sin(\varphi)\\
        -\cos(\theta) \sin(\theta)\cos(\phi) & -\cos(\theta)\sin(\theta)\sin(\var\phi) & \sin^2(\theta)
    \end{pmatrix}
\end{equation}

\noindent where $\hat{k}=\vec{k}/k$. The far-field amplitude then becomes:
\begin{equation}
    \bvec{F} =  A(k,r_\infty) e^{i\phi(\bvec{r}_1,\vec{k})} \Big[ \underbrace{(p_{1x} + e^{i\phi(\bvec{r}_2 - \bvec{r}_1,\vec{k})} p_{2x})}_{\equiv C_x} \tensorarrow{\mathfrak{g}}\cdot \bvec{e}_x + \underbrace{(p_{1y} + e^{i\phi(\bvec{r}_2 - \bvec{r}_1,\vec{k})} p_{2y})}_{\equiv C_y} \tensorarrow{\mathfrak{g}}\cdot \bvec{e}_y \Big]
\end{equation}

\noindent where for brevity, we have defined the $C_x, C_y$ coefficients. In order to compute $S_3$, we need to express $\vec{F}$ in the spherical basis. To do so, we introduce the transition matrix:
\begin{equation}
    \tensorarrow{M_{SC}}(\hat{k}) = \begin{pmatrix}
        \sin\theta \cos\varphi& \sin\theta\sin\varphi & \cos\theta \\
        \cos\theta\cos\varphi& \cos\theta\sin\varphi & -\sin\theta \\
        - \sin\varphi& \cos\varphi & 0
    \end{pmatrix}
\end{equation}

\noindent that acts on the vectors as:
\begin{equation}
    \begin{pmatrix} a_r\\ a_\theta \\ a_\varphi \end{pmatrix} = \tensorarrow{M_{SC}} \begin{pmatrix} a_x\\ a_y \\ a_z \end{pmatrix}
\end{equation}

\noindent We can now define the new dyadics: 
\begin{equation}
    \tensorarrow{\mathcal{G}} = \tensorarrow{M_{SC}}\tensorarrow{\mathfrak{g}}
\end{equation}

\noindent so that, in spherical coordinates, the far-field amplitude reads:
\begin{equation}
    \vec{F} = A(k,r_\infty) \phi(\bvec{r}_1) \Big[ C_x \tensorarrow{\mathcal{G}} \cdot\bvec{e}_x + C_y \tensorarrow{\mathcal{G}} \cdot\bvec{e}_y \Big]
\end{equation}

\noindent We can now compute the fourth component of the Stokes vector $S_3$ along a given direction $\hat{k}$ in the far-field $S_3 = |F_+|^2 - |F_-|^2 = 2 \mathfrak{Im}(F_\theta F_\varphi^*)$. We obtain :
\begin{equation}\label{eq:ch4:first_dichroic_analytifcal_expression_developped}
    S_3 = 2|A(k,r_\infty)|^2 \Im\{C_x C_y^*\} \Big[\mathcal{G}_{\theta x} \mathcal{G}_{\varphi y} -  \mathcal{G}_{\theta y} \mathcal{G}_{\varphi x}\Big] = 2|A(k,r_\infty)|^2 \Im\{C_x C_y^*\} \cos \theta
\end{equation}
After some algebra, it simplifies to:
\begin{equation}\boxed{
    S_3 = 2|A(k,r_\infty)|^2 \Im\Big\{ p_1 p_2^* e^{-i\phi^*(\vec{r}_2 - \vec{r}_1;\vec{k})}  \Big\} \cos(\theta) \; \textcolor{red}{\sin(\alpha)}
    }
    \label{eq:DT}
\end{equation}

\noindent We stress the apparition of the $\sin(\alpha)$ factor which will play a major role in the rest of the document. As a reminder:
\begin{equation}
    \phi^*(\vec{r}_2 - \vec{r}_1;\vec{k})=e^{i\vec{k}.(\vec{r}_2-\vec{r}_1)}
\end{equation}

\noindent and $\theta$ corresponds to the polar angle between the $z$-axis and the $\hat{k}$.

\subsection{$p_1 p_2^*$ as a function of the exciting electrical field}

\noindent In the previous section, we have shown that:
\begin{eqnarray}
    p_1 &=& v E_{1}  + w  E_{2}\\
    p_2 &=& w E_{1} + v E_{2}
\end{eqnarray}

\noindent with:
\begin{equation}
v = T^2 \Omega^2, \quad w = -T^2 \omega_c^2
\end{equation}

\noindent Therefore, we can compute the product:
\begin{eqnarray}
    p_1 p_2^* &=& (v E_{1} + w E_{2})(w^* E_{1}^* + v^* E_{2}^*)\\
&=& v w^* E_{1} E_{1}^* + w v^* E_{2} E_{2}^* + w w^* E_{2} E_{1}^* + v v^* E_{1} E_{2}^*\\
&=& v w^* |E_{1}|^2 + v^* w |E_{2}|^2 + |w|^2 E_{2} E_{1}^* + |v|^2 E_{1} E_{2}^*\label{eq:pxpy*}
\end{eqnarray}

\noindent We note that the source for the appearance of dichroism in equation \eqref{eq:DT} comes from the imaginary part of a product involving $p_1 p_2^*$ and the phase term. Ignoring the phase term for simplicity, dichroism can arise either through a circularly polarised excitation (in which case the two first terms in the right hand side of equation \eqref{eq:pxpy*} cancel and the last two terms  remain), or through a linearly polarized excitation (in which case the two last terms of the right hand side of equation \eqref{eq:pxpy*} cancel and one of the two other terms  remains). In the first case, the nature (real or complex) of the coefficient is unimportant, in the second case, only complex terms allow for observing dichroism if the phase term is null. Adding the phase term makes the situation more complicated, as explained later.

\subsection{Circular dichroism cathodoluminescence}

Here we will present the calculation of the far-field dichroism in the case where the incident electrical field is induced by a traveling electron, i.e, the case of circularly polarised cathodoluminescence. We will exemplify the calculation at high symmetry points, i.e  considering the cases where the beam is at one or the other of the tips of the BKS. The case where the beam is in the gap is not covered here because experimentally the data are less reliable in the gap, see main text.
\subsubsection{Calculation of the dichroism when the excitation field stems from an electron beam}
The basic idea is that we will consider that positioning a free electron beam close to the tip of a nanoantenna will generate a constant field along the nanoantenna (therefore its corresponding dipole in the current approximation), which will excite \textit{directly} only this antenna (or dipole in the current approximation). This is conceptually a robust approximation given the properties of electromagnetic fields associated with fast electrons \cite{GarciaDeAbajo2010}. Indeed, the near field of the electron beam possess a spectrum that is relatively white in energy, meaning all resonances of the BKS is likely to be excited. Also, the exciting electrical field from the electron has a cylindrical symmetry along the electron path. Therefore, if the beam is crossing the antenna axis direction, it will efficiently excite the longitudinal modes of the antenna, of interest here, with a field polarized along the given antenna direction. Finally, this exciting field is decreasing quasi exponentially as a function of the distance between the beam to the antenna. The exciting field is therefore not constant along the nanoantenna, but will resonantly drive the antenna longitudinal plasmons in the same way a constant field would. Finally, since the field exponentially decreases, positioning the beam at the external tip of a given antenna (or even the gap tip if the two antenna are sufficiently separated, see for example the case of the 270$^\circ$ BKS in the main text) will not measurably excite the other antenna.\\

\noindent This said, let's consider the case of a BKS represented by the above described two coupled dipoles model. An electron travels along the $z$ direction. The two dipoles are oriented perpendicular to the electron direction. 
Positioning the electron beam close to the tip of the first antenna will be represented by the following exciting field configuration:
\begin{equation} \label{EQ:tip_1_excitation}
E_{1} = E_0, \quad E_{2} = 0
\end{equation}

\noindent where $E_0$ is an arbitrary field amplitude. In the same way, putting the electron at the end of rod 2 corresponds to the following parameters:
\begin{equation} \label{EQ:tip_2_excitation}
E_{1} = 0, \quad E_{2} = E_0
\end{equation}

\noindent To simplify the readability of the algebraic developments in the following sections, we introduce the reduced $S_3$ Stokes vector as:
\begin{equation} \label{EQ:reduced_S3_vector}
S_{3r} = \frac{S_3}{2 |A(k, r_\infty)|^2}
\end{equation}

\noindent Combining equations \eqref{eq:DT}, \eqref{eq:pxpy*}, \eqref{EQ:tip_1_excitation} and \eqref{EQ:reduced_S3_vector}:
\begin{equation}
S_{3r}(1) = \Im(v w^* E_0^2 e^{i \vec{k} \cdot (\vec{r}_1 - \vec{r}_2)}) \cos(\theta)\sin(\alpha)
\end{equation}

\noindent where the "1" in $S_{3r}(1)$ indicates that the electron beam is positioned to excite the dipole $1$ and:
\begin{equation}
v w^* = -T^2 \Omega^2 T^{*2} \omega_c^2 = -|T^2|^2 \Omega^2 \omega_c^2
\end{equation}

\noindent Combining the two previous equations give the final result:
\begin{equation}\label{EQ:S3r_tip1}
S_{3r}(1) = -E_0^2 |T^2|^2 \omega_c^2 \; \Im(\Omega^2  e^{i \vec{k} \cdot (\vec{r}_1 - \vec{r}_2)}) \cos(\theta)\sin(\alpha)
\end{equation}

\noindent we recall that $\Omega^2=\omega_0^2 - i\gamma\omega - \omega^2 \in \mathbb{C}$. Similarly, if the electron excites the second rod, we obtain:
\begin{equation} \label{EQ:S3r_tip2}
S_{3r}(2) = -E_0^2 |T^2|^2 \omega_c^2 \; \Im((\Omega^*)^2  e^{i \vec{k} \cdot (\vec{r}_1 - \vec{r}_2)}) \cos(\theta)\sin(\alpha)
\end{equation}

\subsubsection{First-order development of the dichroism}

We now assume the distance between two antennas is much smaller than the wavelength of light. By doing so, we also implicitly assume that each nano-antenna is small with respect to the wavelength of light. Then:

\begin{equation}
|\vec{k} \cdot (\vec{r}_1 - \vec{r}_2)| \ll 1
\end{equation}

\noindent Using this assumption, we apply a Taylor expansion of the phase term to the first order i.e.:
\begin{equation}
    \exp(i \vec{k} \cdot (\vec{r}_1 - \vec{r}_2)) \approx 1+i\vec{k}.(\vec{r}_1-\vec{r}_2)
\end{equation}

\noindent which leads to :
\begin{equation}
S_{3r} (1) \approx -E_0^2 |T^2|^2 \omega_c^2  \Im(\Omega^2(1 + i\vec{k} \cdot (\vec{r}_1 - \vec{r}_2))) \cos(\theta) \sin(\alpha)
\end{equation}

\noindent which can be re-written as:
\begin{equation}
\boxed{
S_{3r}(1) = -E_0^2 |T^2|^2 \omega_c^2  \Big[ \textcolor{red}{\underbrace{\Im(\Omega^2)}_{\text{Order 0}}} + \textcolor{blue}{\underbrace{\Re(\Omega^2) \vec{k} \cdot (\vec{r}_1-\vec{r}_2)}_{\text{Order 1}}} \Big] \cos(\theta)\sin(\alpha)
}
\end{equation}

\noindent Applying the same procedure to \eqref{EQ:S3r_tip2}, we obtain
\begin{equation}
\boxed{
S_{3r}(2) = -E_0^2 |T^2|^2 \omega_c^2  \Big[ \textcolor{red}{\underbrace{-\Im(\Omega^2)}_{\text{Order 0}}} + \textcolor{blue}{\underbrace{\Re(\Omega^2) \vec{k} \cdot (\vec{r}_1-\vec{r}_2)}_{\text{Order 1}}} \Big] \cos(\theta)\sin(\alpha)
}
\end{equation}

\noindent where we have used $\Im((\Omega^*)^2)=-\Im(\Omega^2)$ and $\Re((\Omega^*)^2)=\Re(\Omega^2)$. Crucially, when switching from exciting dipole 1 to dipole 2, the sign of the order 0 term is flipped while order 1 remains unchanged. 
 
\subsubsection{Dichroic signal at order 0}

\noindent Since $\Im(\Omega^2)=-\gamma\omega$, the zero order terms (labeled with the superscript $(0)$) read:
\begin{equation}
    S^{(0)}_{3r}(1) = -S^{(0)}_{3r}(2) =E_0^2 |T^2|^2 \omega_c^2  \gamma \omega \cos(\theta) \sin(\alpha)
    \label{eq:S3_0}
\end{equation} 

\subsubsection{Dichroic signal at order 1}
Similarly, since $\Re(\Omega^2)= \omega_0^2-\omega^2$ and $\vec{k}.(\vec{r}_1-\vec{r}_2)=k_x a-k_y a -k_z d$, the first order terms (labeled with the superscript $(1)$) read:
\begin{equation}
    S^{(1)}_{3r}(1) = S^{(1)}_{3r}(2) = -E_0^2 |T^2|^2 \omega_c^2 (\omega_0^2-\omega^2)\left[ k_x a - k_y a - k_z d \right]) \cos(\theta) sin(\alpha)
\end{equation}

\subsubsection{Sum and difference of the signal acquired at each tips}

\noindent Summing the two signals (respectively acquired at tip $1$ and $2$) will therefore cancel the zero order terms i.e.:
\begin{equation}
\boxed{S_{3r}(1) + S_{3r}(2) = -2 E_0^2 |T^2|^2 \omega_c^2 (\omega_0^2-\omega^2)\left[ k_x a - k_y a - k_z d \right]) \cos(\theta) sin(\alpha)}
\end{equation}

\noindent Similarly, taking the difference between the two terms will only retain the zero order terms:
\begin{equation}
\boxed{S_{3r}(1) - S_{3r}(2) = 2E_0^2 |T^2|^2 \omega_c^2  \gamma \omega \cos(\theta) \sin(\alpha)}
\end{equation}

\subsection{Normalisation of dichroism by the integral of the collected intensity} \label{SEC:full_normalization}
\noindent To compare with our experiment, we need to compute the $S_3$ parameters normalized to the total emitted power that we note $\gsca= S_3/S_0$. The intensity collected along a given direction $\hat{k}$ reads:
\begin{equation}
\begin{split}
    S_0 = |F_\theta|^2 + |F_\varphi|^2\\ = |A(k,r_\infty)|^2 \Big[ & (|p_x|^2 + |p_y|^2) \frac{3+\cos(2\theta)}{4} + (|p_x|^2 - |p_y|^2)\frac{\cos(2\varphi)(\cos(2\theta)-1)}{4} \\ & - 2\cos(\varphi)\sin(\varphi)\sin^2(\theta)  \Re\{ p_x p_y^* \phi^*(\vec{r}_2-\vec{r}_1, \hat{k}) \} \Big]
\end{split}
\end{equation}

\noindent making the same first order development of the phase term, we simplify the real part as: 
\begin{equation}
\begin{split}    
    \Re\{ p_x p_y^* \phi^*(\vec{r}_2-\vec{r}_1, \hat{k}) \} & \simeq -\omega_c^2 |E_0|^2 |T^2|^2 \Re\{ (\omega_0^2 - \omega^2 -i \gamma \omega) (1 + i\bvec{k} \cdot (\bvec{r}_2 - \bvec{r}_1)) \} \\
    & \simeq -\omega_c^2 |E_0|^2 |T^2|^2 \Big[ \omega^2_0 - \omega^2 + \gamma \omega \bvec{k}\cdot (\bvec{r}_2 - \bvec{r}_1) \Big]
\end{split}
\end{equation}

\noindent In the next sections, we will need to compute the integral of $S_3$ and $S_0$ on the entire or part of the far-field sphere. Here, we start by computing the integral of $S_0$ on the entire far-field sphere i.e. for $4\pi$ steradians i.e. for $\theta\in [0;\pi ]$ and $\varphi\in\{0;2\pi \}$. We then define an intermediate quantity, $\overline{S}_3$ as $S_3$ divided by the $S_0$ integrated on the $4\pi$ or $2\pi$. This intermediate quantity is \textit{not} $g$, but makes the discussion on symmetries in the main paper easier to follow.  Defining:
\begin{equation}
    \mathcal{S}_{0,4\pi} = \oiint_{4\pi} S_0 \;d\Omega
\end{equation}

\noindent The term $$\oiint - 2\cos(\varphi)\sin(\varphi)\sin^2(\theta)  \Re\{ p_x p_y^* \phi^*(\vec{r}_2-\vec{r}_1, \hat{k}) \} d\Omega$$ sums up to 0 when integrating over the full range for $\varphi$. The same occurs for $$\oiint \frac{\cos(2\varphi)(\cos(2\theta)-1)}{4}d\Omega = 0$$ Only remains 
\begin{equation}
    \oiint \frac{3+\cos(2\theta)}{4} d\Omega = \frac{8 \pi}{3}
\end{equation}

\noindent which leads to:
\begin{equation}
     \mathcal{S}_{0,4\pi} = \frac{8 \pi}{3} |A(k,r_\infty)|^2 (|p_x|^2 + |p_y|^2) 
\end{equation}

\noindent and:
\begin{equation}
    |p_x|^2 + |p_y|^2 = |E_0|^2 |T^2|^2 ( |\Omega^2|^2 + \omega_c^4) 
\end{equation}

\noindent Combining the two previous equations, we obtain:
\begin{equation}
     \mathcal{S}_{0,4\pi} = \frac{8 \pi}{3} |A(k,r_\infty)|^2 |E_0|^2 |T^2|^2 ( |\Omega^2|^2 + \omega_c^4) 
\end{equation}

\noindent One could show that the signal integrated over the upper (equivalently lower) far-field hemisphere only is exactly half this value:
\begin{equation}
\boxed{
     \mathcal{S}_{0,2\pi} = \frac{1}{2} \mathcal{S}_{0,4\pi} = \frac{4 \pi}{3} |A(k,r_\infty)|^2 |E_0|^2 |T^2|^2 ( |\Omega^2|^2 + \omega_c^4) 
     }
\end{equation}

\noindent Besides, we have shown in the previous section that:
\begin{equation}
    S_3(1) = - 2 |A(k, r_\infty)|^2 E_0^2 |T^2|^2 \omega_c^2 \; \Im(\Omega^2  e^{i \vec{k} \cdot (\vec{r}_1 - \vec{r}_2)}) \cos(\theta)\sin(\alpha)
\end{equation}

\noindent Therefore:
\begin{equation}
    \overline{S}_3(1) \equiv \frac{S_3(1)}{\mathcal{S}_{0,4\pi}} = \frac{3}{4\pi} \frac{-\omega_c^2}{|\Omega^2|^2 + \omega_c^4} \Im\{\Omega^2 e^{i \bvec{k}\cdot(\bvec{r}_1 - \bvec{r}_2)}\} \cos (\theta) \sin(\alpha)
\end{equation}

\noindent and, after the first order expansion in $\vec{k}.(\vec{r}_1-\vec{r}_2)$:
\begin{equation}
        \overline{S}_3(1) = \frac{S_3(1)}{\mathcal{S}_{0,4\pi}} = \frac{3}{4\pi} \frac{\omega_c^2}{|\Omega^2|^2 + \omega_c^4} \Big[ \gamma \omega + (\omega^2_0 - \omega^2) \bvec{k}\cdot(\bvec{r}_1 - \bvec{r}_2) \Big] \cos(\theta) \sin(\alpha)
\end{equation}

\noindent and similarly for the electron beam positioned at tip 2:
\begin{equation}
        \overline{S}_3(2) = \frac{S_3(2)}{\mathcal{S}_{0,4\pi}} = \frac{3}{4\pi} \frac{\omega_c^2}{|\Omega^2|^2 + \omega_c^4} \Big[ - \gamma \omega + (\omega^2_0 - \omega^2) \bvec{k}\cdot(\bvec{r}_1 - \bvec{r}_2) \Big] \cos(\theta) \sin(\alpha)
\end{equation}

\subsection{Consequences of the use of a finite size mirror}\label{subsec:consequence}

In a real detection set-up, the mirror will most likely be placed in a half-space separated by the mid-plane of the BKS structure.   We therefore now turn to the question of applying the proposed method with a mirror still located in one of these half-spaces, but  covering a solid angle smaller than $2\pi$. 

\subsubsection{Zeroth order term}

The zeroth order term in $S_3$ is such that (see equation \ref{eq:S3_0}):
\begin{equation}
S^{(0)}_3(1)+S^{(0)}_3(2)  = 0 
\qquad \forall \theta,\; \forall \varphi
\end{equation}

\noindent and therefore is zero whatever the form of the detector. This means that the most of the ``hidden chirality'' (see main text and \cite{Zu2018}) is entirely washed out by the proposed symmetrization, whatever the solid angle. We won't discuss this term anymore.

\subsubsection{First order term}

The first order term is split in two terms:

\begin{itemize}
\item in-plane term:
\begin{equation}
S_3^{(1),\perp} = S_3^{(1),\perp}(1) = S_3^{(1),\perp}(2) \propto \sin\theta \, (\Delta x \cos\varphi + \Delta y \sin\varphi)\, \sin\alpha
\end{equation}

where $\Delta x = -asin(\alpha)$ ($\Delta y = a(1-cos(\alpha)$) is the difference in dipole position along $x$ ($y)$

\item out-of-plane term:
\begin{equation}
S_3^{||} \propto \cos\theta \, \sin\alpha \, d
\end{equation}
\end{itemize}

\bigskip

\noindent If the detecting mirror is a cone centered on the $z$ direction, by symmetry
($\theta \in [0,\theta_0]$ and $\varphi \in [0,2\pi]$):
the in-plane term is null
\begin{equation}
\int_{\Omega} S_3^{(1),p} = 0
\end{equation}

\noindent while the out-of-plane is just renormalized by the sine of the detected cone:
\begin{equation}
\int_{\Omega} S_3^{\perp} \propto d \sin\alpha \, \sin\theta_0
\end{equation}

\noindent therefore, for a cone detection, $g$ has the same energy ($\omega$), gap ($d$)  and angle ($\alpha$) dependence as if it would be integrated on $2\pi$, but the absolute value might be modified, as the numerator and denominator ($S_3$ et $S_0$) may have slightly different angular dependence.

\bigskip

\noindent If now the revolution symmetry is broken ($\phi \in [\phi_a,\phi_b]$), the out-of plane contribution reads:
\begin{equation}
\int_{\Omega} S_3^{||} \propto d \sin\alpha \, \cos\theta_0 \, \Delta\varphi
\end{equation}

 \noindent with $\Delta\varphi = \phi_a -\phi_b$.
 Again, if we were to consider only this part, \gsca would only be slightly renormalized, with the same  energy ($\omega$), $d$ and $\alpha$ dependence) as if it would be integrated on $2\pi$ (or $4\pi$).

\noindent However, the in-plane component is more complicated:
\begin{equation}
\int_{\Omega} S_3^{\perp} \propto sin^3(\theta_0)[-sin(\alpha)(sin(\phi_a)-sin(\phi_b))+(1-cos(\alpha))(cos(\phi_a)-cos(\phi_b))]sin(\alpha)
\end{equation}

\noindent We see that this term adds extra dependence in $\alpha$, typically adding a second harmonic (oscillations in $2\alpha$). It acts as another source of "hidden chirality", to the best of our knowledge not previously discussed in \cite{Zu2018}. We note that proper symmetrization of the BKS orientation with respect to the mirror may nullify this term. This would be the case for an off-axis parabolic mirror which revolution axis being orientated along the direction of the bisector of the two dipoles direction. However, such a systematic orientation of the BKS with respect to the mirror would be in practice very laborious.

\bigskip

\noindent It means that if the detector is reasonably revolution-symmetric, the actual measured
$g_{\mathrm{sca}}$ is only slightly re-normalized.

As the geometry of an off-axis parabolic mirror is not easy to reproduce, we have made
some simulations with the mirror geometry compared to a half-sphere, described in section \ref{sec:real_mirror}. We note however that the absence of harmonic in the experimental data already  points out an absence of relevant effect of the finite size of the experimental mirror on the conclusion of the main text.

\subsection{Normalization of the integrated dichroism signal by the total detected intensity}

In our experiment, we only detect the light emitted on a single hemisphere of the far-field sphere (see main text). We therefore need to compute the $g$ but only on one hemisphere. From our calculations developed in section \ref{SEC:full_normalization}, we have:
\begin{equation}
        \gsca (1) = \frac{S_3(1)}{\overline{S}_{0,2\pi}} = \frac{3}{2\pi} \frac{\omega_c^2}{|\Omega^2|^2 + \omega_c^4} \Big[ \gamma \omega + (\omega^2_0 - \omega^2) \bvec{k}\cdot(\bvec{r}_1 - \bvec{r}_2) \Big] \cos(\theta) \sin(\alpha)
\end{equation}

\noindent and:
\begin{equation}
        \gsca (2) = \frac{S_3(2)}{\overline{S}_{0,2\pi}} = \frac{3}{2\pi} \frac{\omega_c^2}{|\Omega^2|^2 + \omega_c^4} \Big[ - \gamma \omega + (\omega^2_0 - \omega^2) \bvec{k}\cdot(\bvec{r}_1 - \bvec{r}_2) \Big] \cos(\theta) \sin(\alpha)
\end{equation}

\noindent \textbf{Please note that the normalization factor is now $\overline{S}_{0,2\pi}$}.\\

\noindent Let's now compute:
\begin{equation}
    \gsca_{2\pi} (1) = \int_0^{\pi/2} \sin{\theta} d\theta \int_0^{2\pi}d \varphi \;  \gsca (1)
\end{equation}

\noindent We start with the zero order term:
\begin{eqnarray}
    \gsca^{(0)}_{2\pi} (1) &=& \frac{3}{2\pi} \frac{\omega_c^2}{|\Omega^2|^2 + \omega_c^4} \gamma\omega \sin(\alpha) \int_0^{\pi/2} \sin{\theta} d\theta \int_0^{\pi}d \varphi \;  \cos(\theta)\\
    &=& 3 \frac{\omega_c^2}{|\Omega^2|^2 + \omega_c^4} \gamma\omega \sin(\alpha) \underbrace{\int_0^{\pi/2} \sin{\theta} \cos(\theta) d\theta }_{=1/2}\\
    &=& \frac{3}{2} \frac{\omega_c^2 \gamma\omega}{|\Omega^2|^2 + \omega_c^4}  \sin(\alpha)
\end{eqnarray}

\noindent Besides, the term $\vec{k}.(\vec{r}_1-\vec{r}_2)$ will give:
\begin{equation}
    \vec{k}.(\vec{r}_1-\vec{r}_2) = k\sin(\theta)\cos(\varphi)(x_1-x_2)+k\sin(\theta)\cos(\varphi)(y_1-y_2)+ k\cos(\theta)\underbrace{(z_1-z_2)}_{=-d}
\end{equation}

\noindent The two first terms will integrate to 0 over $\varphi$ so only the last term (in $z$) can contribute to the first order term:
\begin{eqnarray}
    \gsca^{(1)}_{2\pi} (1) &=& \frac{3}{2\pi} \frac{\omega_c^2 (\omega_0-\omega^2)}{|\Omega^2|^2 + \omega_c^4}  \sin(\alpha) \int_0^{\pi/2} \sin{\theta} d\theta \int_0^{\pi}d \varphi \;  (-dk\cos(\theta))\cos(\theta)\\
    &=&  -3 \frac{\omega_c^2 (\omega_0-\omega^2)}{|\Omega^2|^2 + \omega_c^4}  k d\sin(\alpha) \underbrace{\int_0^{\pi/2} \sin{\theta} \cos^2(\theta) d\theta}_{=1/3}\\
    &=& -\frac{\omega_c^2 (\omega_0-\omega^2)}{|\Omega^2|^2 + \omega_c^4}  k d\sin(\alpha) 
\end{eqnarray}

\noindent Collecting the two terms we finally obtain:
\begin{eqnarray}
    \gsca_{2\pi} (1) &=& \gsca^{(0)}_{2\pi}(1)+\gsca^{(1)}_{2\pi}(1)\\
    &=& \frac{3}{2} \frac{\omega_c^2 \gamma\omega}{|\Omega^2|^2 + \omega_c^4}  \sin(\alpha) -\frac{\omega_c^2 (\omega_0-\omega^2)}{|\Omega^2|^2 + \omega_c^4}  k d\sin(\alpha)   
\end{eqnarray}

\noindent Similarly:
\begin{eqnarray}
    \gsca_{2\pi} (2) &=& \gsca^{(0)}_{2\pi}(2)+\gsca^{(1)}_{2\pi}(2)\\
    &=& -\frac{3}{2} \frac{\omega_c^2 \gamma\omega}{|\Omega^2|^2 + \omega_c^4}  \sin(\alpha) -\frac{\omega_c^2 (\omega_0-\omega^2)}{|\Omega^2|^2 + \omega_c^4}  k d\sin(\alpha)   
\end{eqnarray}

\noindent In particular, we obtain the important result:
\begin{equation}\label{EQ:g_order1_integrated_pi}
\boxed{
    \gsca_{2\pi} (1)+\gsca_{2\pi} (2) = -2\frac{\omega_c^2 (\omega_0-\omega^2)}{|\Omega^2|^2 + \omega_c^4}  k d\sin(\alpha)   
    }
\end{equation}

\subsection{Relation between the normalized dichroic signal and the plasmonic helicity}

\noindent We will now compute the time-averaged helicity of the BKS modes using formula \eqref{EQ:harmonic_plasmon_helicity}. We consider the case where the electron impinges a tip of the BKS (for example tip 1):
\begin{eqnarray}
    \vec{p}_1 &= E_0T^2 \Omega^2 \vec{e}_1\\
    \vec{p}_2 &= E_0T^2 \omega_c^2 \vec{e}_2
\end{eqnarray}

\noindent Therefore:
\begin{equation}
    \vec{p}_1 \times \vec{p}^*_2 = - E_0^2 \vert T^2 \vert^2 \Omega^2 \omega_c^2 \sin(\alpha) e_z
\end{equation}

\noindent where the minus sign comes from the cross-product $\vec{e}_1\times \vec{e}_2$. Therefore:
\begin{equation}
    (\vec{p}_1 \times \vec{p}^*_2).(\vec{r}_2-\vec{r}_1) = - E_0^2 \vert T^2 \vert^2 \Omega^2 \omega_c^2 d\sin(\alpha)
\end{equation}

\noindent We can now take the real part of the previous equation. The only complex factor is $\Omega^2$ (see \eqref{EQ:def_Omega_squared}), therefore:
\begin{equation}
    \Re \{(\vec{p}_1 \times \vec{p}^*_2).(\vec{r}_2-\vec{r}_1) \} =  - E_0^2 \vert T^2 \vert^2 (\omega_0^2-\omega^2) \omega_c^2 d\sin(\alpha)
\end{equation}

\noindent Plugging the latter in \eqref{EQ:harmonic_plasmon_helicity}, the helicity of the BKS system is therefore given by:
\begin{equation}
    \Held(\alpha,\omega) =  - \frac{\aleph}{2} E_0^2 \vert T^2 \vert^2 (\omega_0^2-\omega^2) \omega_c^2 d\sin(\alpha)
\end{equation}

\noindent \textbf{where the superscript "d" has been added to stress that it corresponds to the definition of our plasmonic helicity applied to the two dipoles model. We also added the $\alpha$-dependence explicitly, as it is a parameter that we can vary in our experiment}. Since all our experiments are normalized to the total energy emitted by the system within the experimentally detection solid angle $\Omega_{exp}$ (see main text), for comparison to experiment, we need the dipole-amplitude normalized helicity $\Heldnorm$ defined as:
\begin{eqnarray}
    \Heldnorm(\alpha,\omega)&=& \frac{\Held(\alpha,\omega)}{\vert p_x\vert^2+\vert p_y\vert^2}\\
    &=& -\frac{\aleph}{2} \frac{\omega_c^2(\omega^2_0- \omega^2)}{|\Omega^2|^2 +\omega_c^4} d \sin(\alpha)
\end{eqnarray}

\noindent Comparing this equation with \eqref{EQ:g_order1_integrated_pi}, it follows that:
\begin{equation}
\boxed{
   \gsca_{2\pi} (1)+\gsca_{2\pi} (2) = \frac{4}{\aleph} \; k \; \Heldnorm(\alpha,\omega)
    }
\end{equation}

\noindent \textbf{which constitutes the main theoretical result of this paper.}

\section{Simulations}

\subsection{pyGDM}
\label{sec:pygdm}

PyGDM is an open-source Python library which aims to perform nano-optic simulations based on the Green Dyadic Method (GDM) by discretizing the volume of  the considered nanoparticles. One can find a complete description of the library in \cite{wiecha2018pygdm, wiecha2022pygdm}. The BKS are modeled as stacked rectangular prisms with dimensions of 150 nm $\times$ 30 nm $\times$ 60 nm. The gap between the antenna is set to 40 nm from the middle of the bottom antenna to the bottom of the top one. The voxel size is set to 10 nm. The antennas are modeled in gold using the Johnson–Christy data and are simulated in vacuum. The regime considered here is the retarded regime. A simulated BKS representation is given in Figure \ref{fig:BKS_sim_pygdm}.

\begin{figure}[h!]
    \centering
    \includegraphics[width=0.8\linewidth]{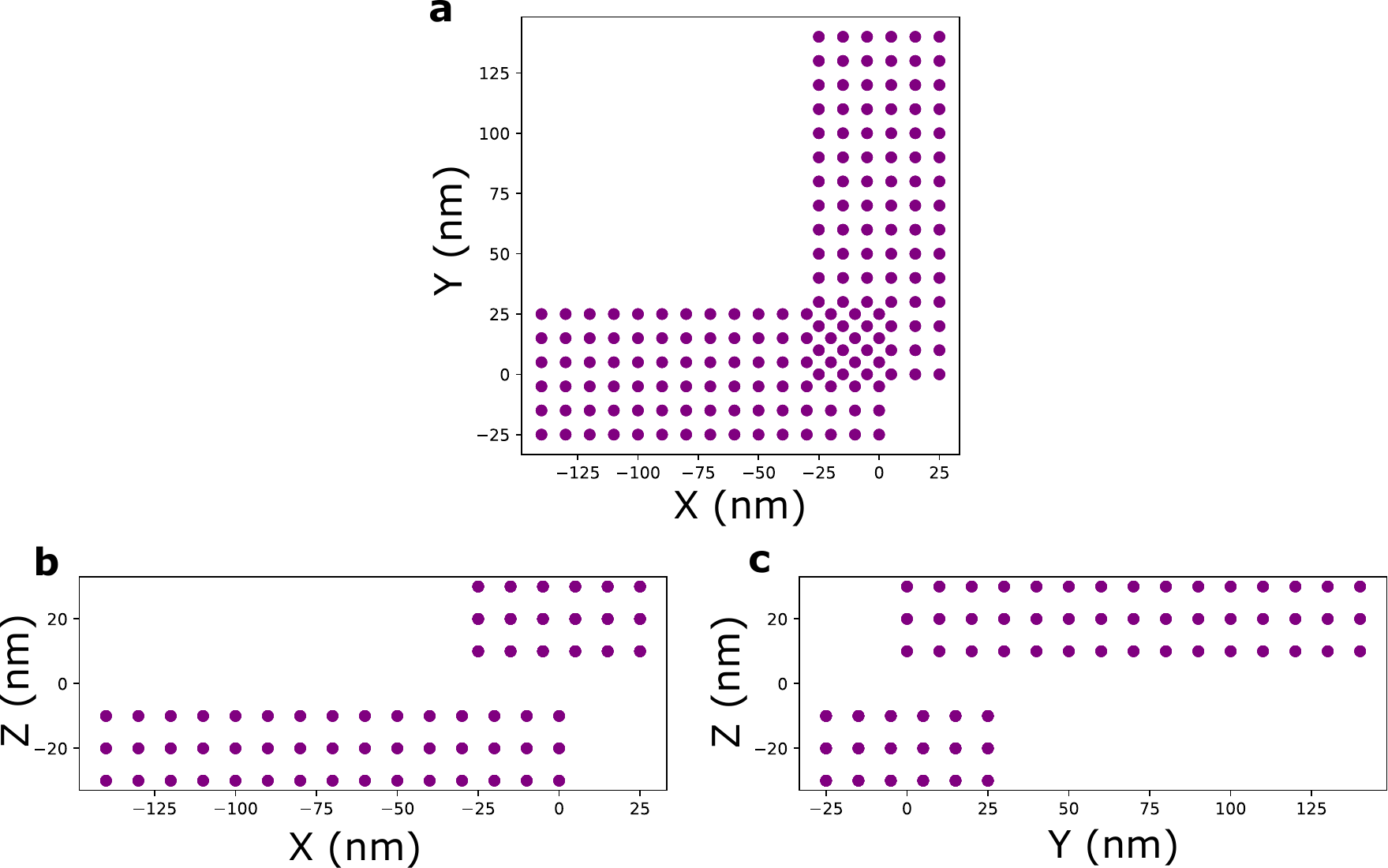}
    \caption{pyGDM BKS representation. (a) Top view of a simulated BKS (b) and (c) side views projected alors the Y and X-axis respectively.}
    \label{fig:BKS_sim_pygdm}
\end{figure}

To simulate all the quantities presented in this article, we had to adapt the PyGDM code by adding homemade developments to distinguish circular polarisations. To calculate the dichroism at a fixed wavelength and a fixed position of the electron beam on the sample, we proceed as follows:

\begin{itemize}
    \item We compute the intensity over the far-field sphere for left circular polarization in each emission direction.
    \item We repeat the previous calculation for right circular polarisation.
    \item We integrate the signals over the portion of the far-field sphere collected by the mirror (in our case, a hemisphere). This yields  $I_L$  and $I_R$, the integrated emission intensities of the left and right polarisations respectively.
    \item We then compare these quantities to determine the circular dichroism at the initially fixed wavelength and position.
\end{itemize}

These steps are repeated for each wavelength and each position on the sample to obtain the spectra and spatial maps.

\subsection{MNPBEM17}

The simulations shown on Fig. 4(d) of the main text were performed using \texttt{MNPBEM 17} toolbox \cite{Hohenester2014}. The two metallic rods were modeled using the Jonhson and Christy dielectric constant for gold. The two rods were meshed with 1083 polygons. The surface charges and currents induced by the electron beam on the rods are calculated using the \texttt{electronbeam} method. Using the \texttt{compgreenret}, we then compute the field generated by these source on the far-field sphere. The latter is modeled by a sphere centered on the BKS system which radius corresponds to 20 times the size of the BKS system. The far-field sphere is meshed using 400 polygons. The electromagnetic field at each polygon is then projected on the circular basis in order to compute the $S_3(\vec{k})$ field. 

\section{Sample Preparation}

 BKS samples compatible with electron microscopy were fabricated using a two-step electron-beam lithography process on a 15 nm thick silicon nitride membrane TEM grid (Ted Pella, Inc.). A standard electron-beam lithography and lift-off process was employed to create the bottom gold nano-antenna, each measuring approximately 150 nm in length, 50 nm in width, and 30 nm in thickness. A 10 nm thick silica spacer was then deposited by thermal evaporation at a 45$^{\circ}$ angle with a rotating sample holder, ensuring conformal coverage of all facets of the gold antenna. The final step consisted of a second electron-beam lithography process, in which alignment marks were used to precisely position the top gold nano-antenna over the bottom one, with the same dimensions. Figure \ref{fig:litho BKS} presents the High Angle Dark Field (HADF) images of 12 BKS with varying angles presented in this study.

\begin{figure}[h!]
    \centering
    \includegraphics[width = \textwidth]{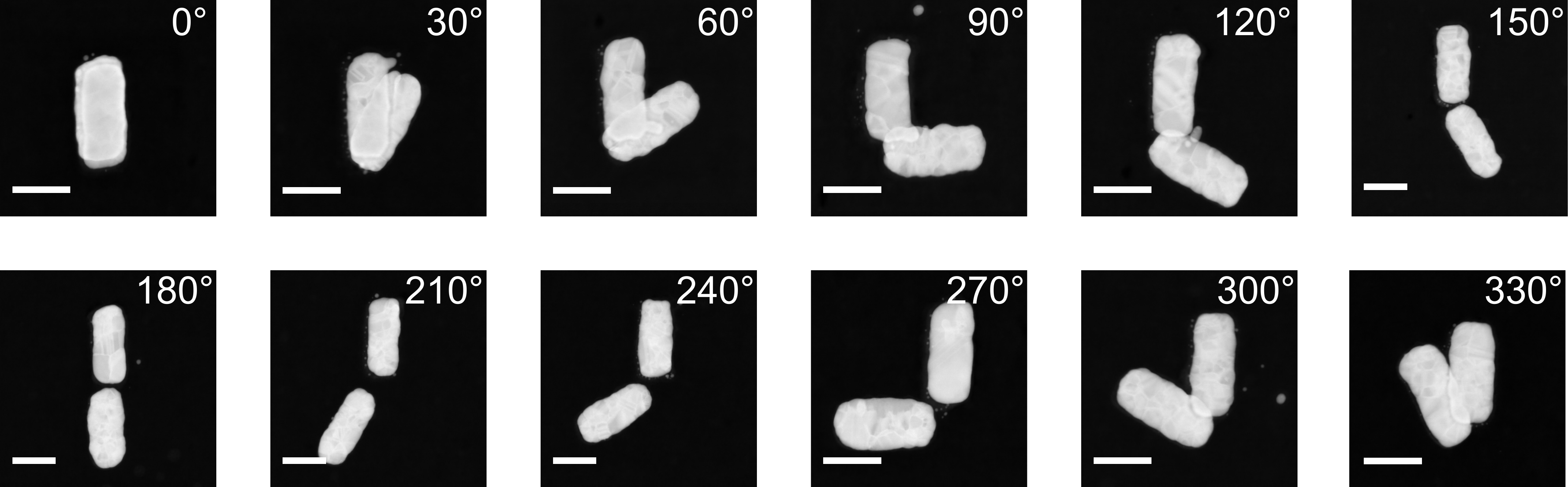}
    \caption{\textbf{HADF images of gold lithographed BKS}.  The angles represented here are going from 0° up to 330° with a step of 30°. The scale bar represents 80 nm.}
    \label{fig:litho BKS}
\end{figure}

Please note the relative position of the antenna is extremely difficult to control and so this drives to misalignment especially at the overlap region, see Figure \ref{fig:litho BKS} BKS 270° and 210° for example.

\newpage

\section{EELS Experimental Setup and Data Analysis}

\subsection{Experimental Setup}

In order to characterize the spectral properties of the BKS, we performed EELS experiments on a NION HERMES-200 with a spectral resolution of 30 meV. Two acceleration voltages were used indistinguishably: 60 kV and 100 kV, with respective incident angles of 25 mrd and 30 mrd. The collection angle was then set to 60 mrd.

Spectral-images were acquired on every BKS structure with a typical dwell time of 50 ms, and a sampling of 7 nm.

\subsection{EELS Results}

Figure \ref{fig:EELS} presents typical EELS spectra  at the tip and the gap of a 90$^{\circ}$ BKS and related energy-filtered EELS images extracted from a spectrum-image. At low energy, two modes are clearly visible, which correspond to the bonding and antibonding mode, as confirmed by the corresponding filtered maps. One can see that the Rabi energy of 350 meV corresponds to a strong coupling of the two dipolar modes of each antenna \cite{Kociak2025}.

\begin{figure}[h!]
    \centering
    \includegraphics[width = 0.8\textwidth]{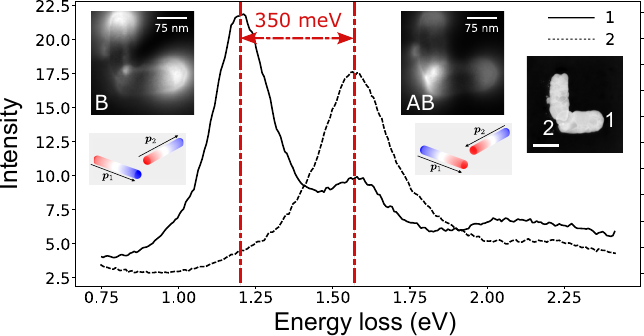}
    \caption{\textbf{Experimental EELS spectra and maps for the 90$^{\circ}$} Spectra taken at the top tip (plain) and gap (dotted) of the BKS. Inset:  maps for the bonding and antibonding modes, related simulated charge densities and corresponding HADF image.}
    \label{fig:EELS}
\end{figure}

We have also systematically extracted the spectra acquired at one tip and at the gap position for every BKS structure studied in this work and compared them with EELS simulations performed using the MNPBEM library as shown in Fig.\ref{fig:BKS_EELS_angle_Diagram}. The two characteristic bonding and anti-bonding modes of the BKS are observed. The energies of these two modes change significantly as we move away from the 0° angle (or 360°). This can be explained by a loss of coupling strength. Indeed, when the angle is $0^{\circ}$, the two antennas completely overlap. However, as the angle increases, the overlap region reduces to the tips of the antennas, becoming minimal at an angle of 180°. The greater the overlap between the antennas, the stronger the coupling, shifting the energies of the two modes accordingly. It can also be noted that the bonding mode is more intense at the tips than at the overlap.. Experimentally, we observe a very similar behavior. 

\begin{figure}[h!]
    \centering
    \includegraphics[width = 0.8\textwidth]{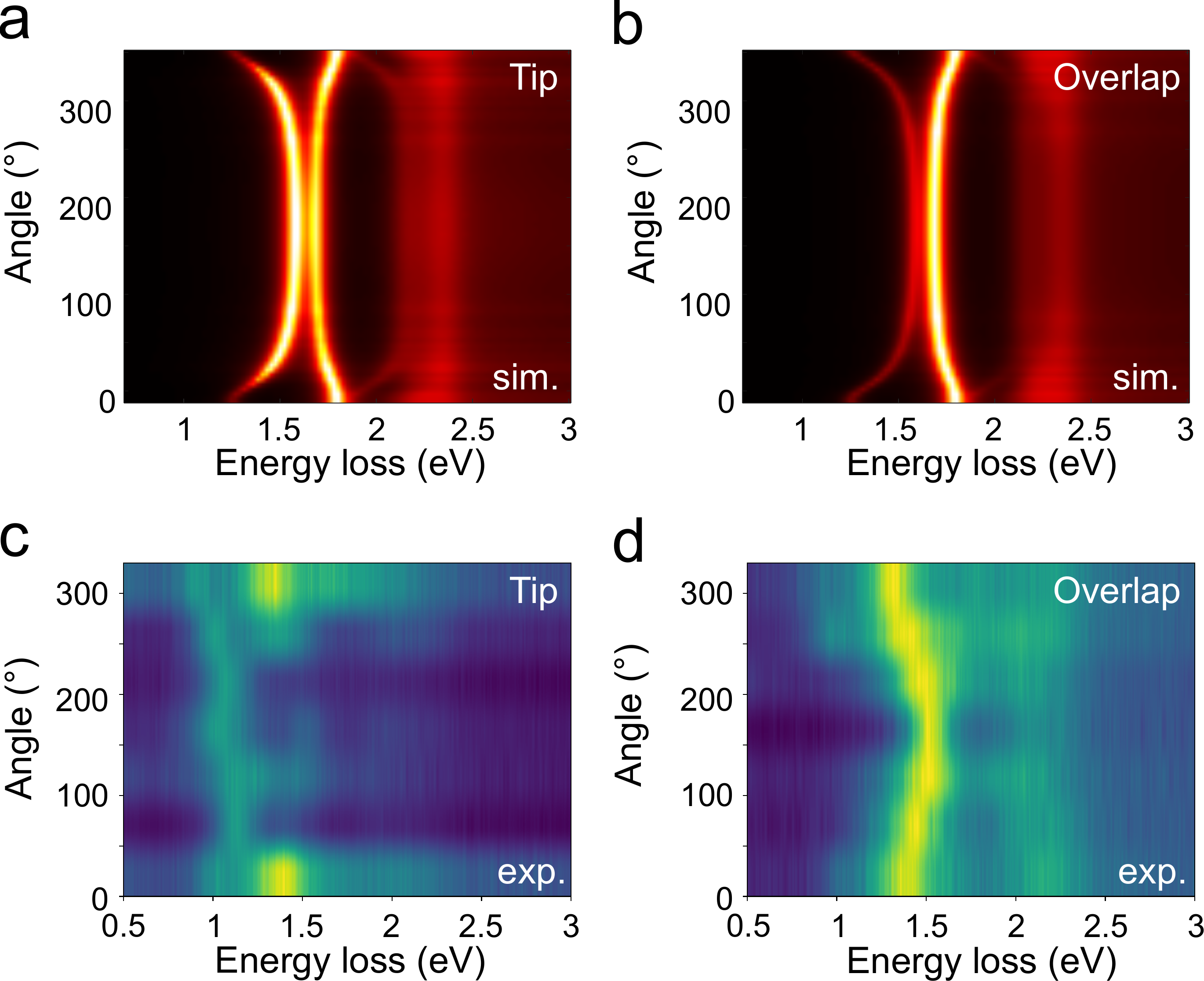}
    \caption{\textbf{Simulated and experimental EELS diagrams of gold BKS with various angles $\alpha$, taken at the tip and at the overlap region.} \textbf{a) and b)} MNPBEM EELS simulations showing the BKS's optical response when the electron beam is positioned at a tip or at the overlap, respectively. \textbf{c) and d)} Experimental EELS diagrams on BKS with angles varying from 0° to 330°.}
    \label{fig:BKS_EELS_angle_Diagram}
\end{figure}

\newpage
\section{pCL Experimental Setup : Parallel Acquisition}

Ideally, the measurements of both polarizations should be carried out under the same conditions. A solution is to ensure that the external variations are consistent at each instant of the dichroism measurement.

 This solution is challenging to implement as it requires acquiring both polarizations simultaneously \cite{Baguenard2023}. However, it offers several advantages. Firstly, the dichroism measurement would take into account the exact same variations for both polarization measurements since they are performed simultaneously. Secondly, since the variations are the same at every moment of the measurement, the acquisition time can be adjusted as needed without worrying about the stability of the machine. Thus, regardless of what happens during the measurement, the signal will remain physical and meaningful.

\begin{figure}[h!]
    \centering
    \includegraphics[width = \textwidth]{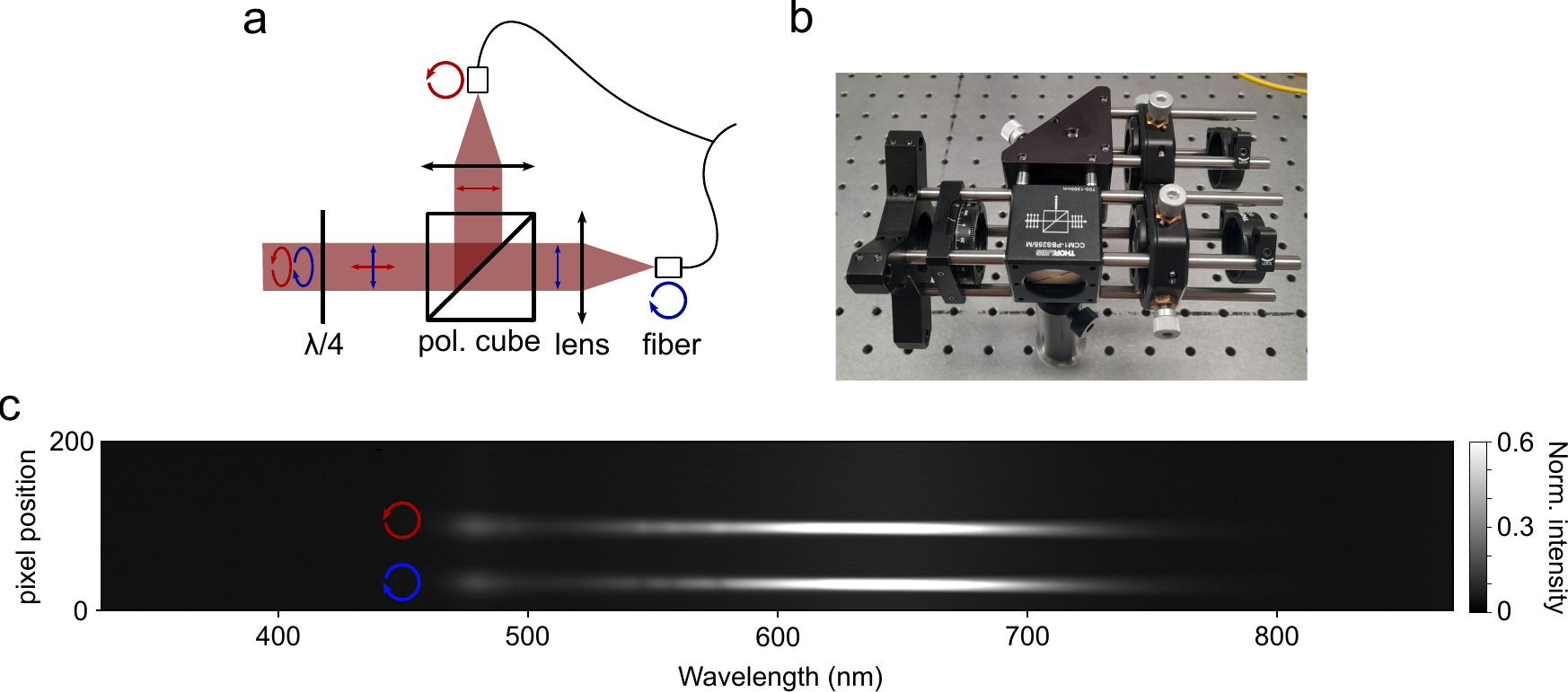}
    \caption{\textbf{Parallel polarization detection pCL setup}. \textbf{a)} Working principle of the polarization parallel acquisition. \textbf{b)} Optical parallel pCL setup used on the NION HERMES-200. \textbf{c)} Optical spectrometer CCD camera image of the two spectra acquired for white light emitted by a LED. The upper one stands for the left-handed polarization, and the lower spectrum is right-handed polarized.}
    \label{1st design}
\end{figure}

A schematic diagram is provided in Fig.\ref{1st design}a. The polarization of the light coming from the microscope can be decomposed into the basis of circular states. The quarter-wave plate allows the transition to the basis of linear polarizations. A polarizing beam splitter spatially separates the right and left circular polarization contributions. Each signal is then collected with a lens and an optical fiber, which is directed to an optical spectrometer. By using a Y-shaped optical fiber, we can simultaneously record two spatially separated spectra on the camera. One corresponding to right circular polarization and the other to left circular polarization as represented in Fig.\ref{1st design}c. The pCL collection setup is shown in Fig.\ref{1st design}b. 

\newpage

\section{Systematic pCL Study on BKS}

Fig.\ref{fig:syst_maps} compiles various maps of BKS at different angles and shows the dichroic signal of the anti-bonding mode summed over 760 nm to 860 nm. To compare the maps, we choose to compare enantiomeric structures, with the exceptions being the achiral structures (BKS at 0° and 180°). BKS with angles below 60° do not show a clear dichroism, likely due to fabrication imperfections. Thus, we will start the analysis with the larger angles. It is observed that BKS with angles of 90°, 120°, and 150° all exhibit the same distribution.

\begin{figure}[h!]
    \centering
    \includegraphics[width = \textwidth]{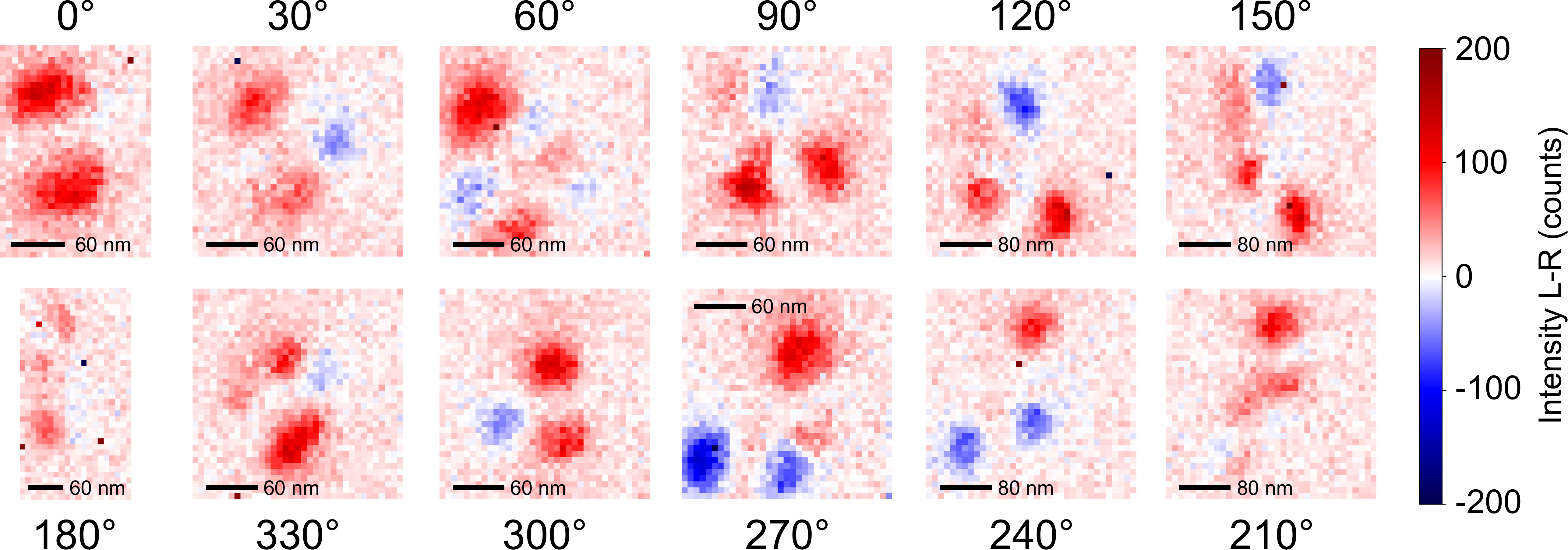}
    \caption{\textbf{Systematic study of circular dichroism pCL maps on various BKS geometry. These spectral images show the anti-bonding mode (760 nm - 860 nm)}. Enantiomers BKS : (30°-330°), (60°-300°), (90°-270°), (120°-240°), and (150°-210°). The 0° and 180° cases, which are achiral structures, are apart.}
    \label{fig:syst_maps}
\end{figure}

Emissions are localized at the tips and the overlap. The polarizations of the tips are opposite, and finally, the polarization of the emissions from the overlap is similar to that of the top antenna. Their enantiomers exhibit exactly the same emission distribution, but the sign of the dichroism is reversed. Additionally, there is a noticeable decrease in the signal intensity as the structures looks like the BKS at 180°. When looking at the maps of BKS with angles less than 60°, the behavior is less clear than for the previous structures. For instance, the local distributions of BKS at 300° and 330° are opposite, whereas we would expect similar maps. More surprisingly, the maps of BKS at 30° and 330° are very similar, which does not correspond to what is observed at larger angles. 

Simulations of these maps were carried out using the Python library pyGDM2 (see Fig. \ref{fig:maps_simulations_pygdm}). They compile all the observed structures and simulate them in vacuum. The details of the simulations are depicted in section \ref{sec:pygdm}. They do not account for the silicon nitride substrate nor the silica layer separating the antennas. The first thing we notice is that there is a good agreement on the localization of the hotspots and their polarization. 

We note that in the case of the experimental 270° BKS (see Fig.\ref{fig:syst_maps}) there is no overlap between the two antennas. This is due to a limit in the precision of our lithography alignment. This configuration is a good opportunity to emphasize that the two polarized signals are localized on two different antennas despite the fact the bounding and anti-bonding plasmons are, per definition, spatially delocalized. This can be straightforwardly understood in the two-dipoles model we have elaborated above. Indeed, \gsca was given for two positions of the electron beam, on either one tip or the other of the BKS, and was shown to have an opposite sign. However, the model is such that the exact position of the electron beam is not relevant, and the important feature is wether one antenna or dipole or the other has been excited. In practice, this is when the electron beam is close to one of the two tips of one of the antenna and further away to any of the two tips of the other antenna. In general, such a condition is not fullfiled in the gap, except in  the very case of the 270° BKS. In this case, we clearly see the rough inversion of \gsca sign for each antenna. Such a behavior is not observed in the other BKS since the overlap hides the effect.

\begin{figure}[h!]
    \centering
    \includegraphics[width = \textwidth]{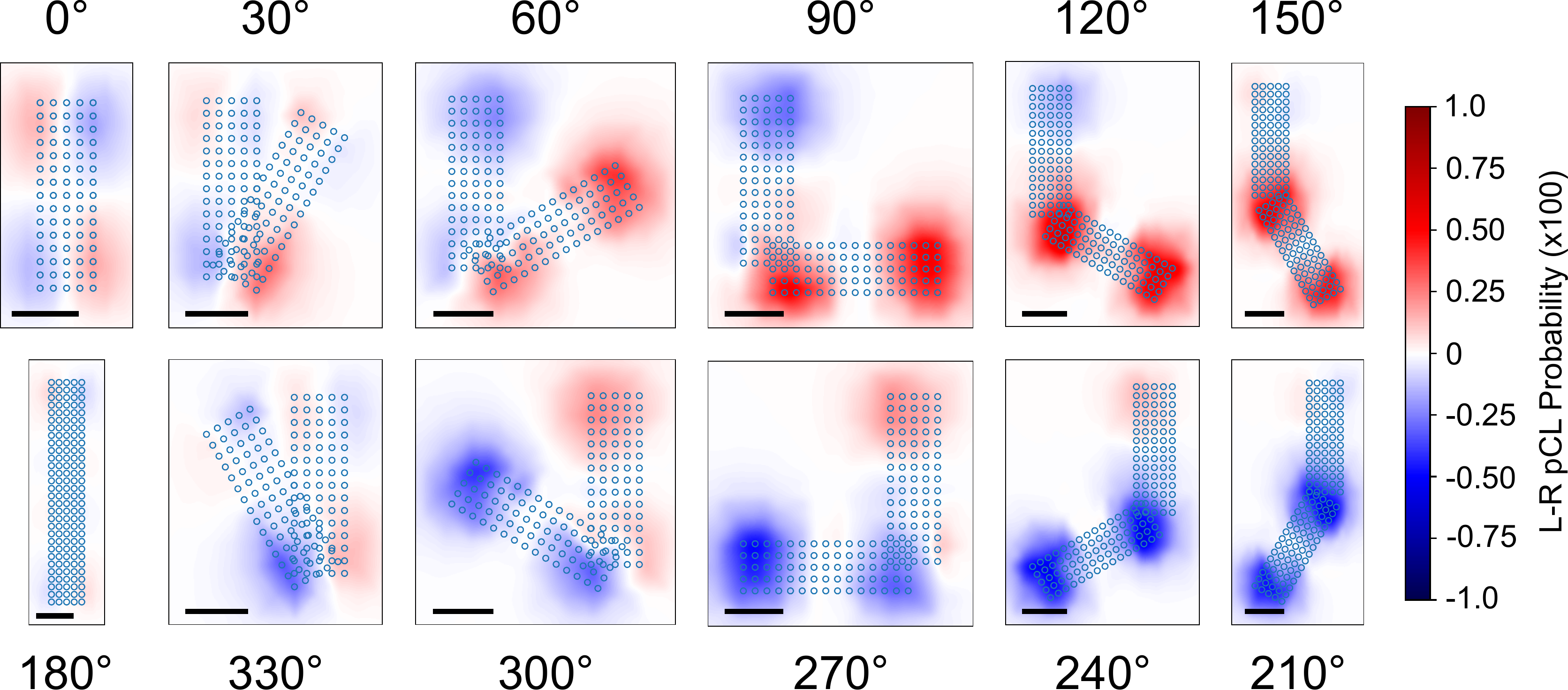}
    \caption{\textbf{Simulated pCL maps for all the BKS observed in Fig.\ref{fig:syst_maps}.} All these maps are filtered at 758 nm which is the anti-bonding mode energy. The signal has been summed over the whole down hemisphere of the far-field. The scale bar represents 50 nm. These simulations were performed using the pyGDM2 python library \cite{arbouet_electron_2014,wiecha_pygdm_2018}.}
    \label{fig:maps_simulations_pygdm}
\end{figure}

\newpage

\section{Simulation Using a Real Mirror Geometry}\label{sec:real_mirror}

In the simulations presented in this paper, the pCL was calculated by integrating the signal over a full hemisphere of the far-field sphere. Experimentally, although it is possible to approach this condition, it is not feasible to collect the entire signal. The simulations that follow therefore take into account the geometry of the mirror used (Attolight Mönch). 

Nevertheless, this mirror collects around 62\% of the hemispherical signal, so considering the hemisphere as a good approximation is not unreasonable. However, the shape of the mirror can introduce an asymmetry in the signal collection depending on its orientation relative to the studied object, see section \ref{subsec:consequence}. This is illustrated in Figure \ref{fig:Dichro90orientationmiroir}, where the dichroic spectra of a 90° BKS are shown for different mirror orientations.

\begin{figure}[h!]
    \centering
    \includegraphics[width=0.7\linewidth]{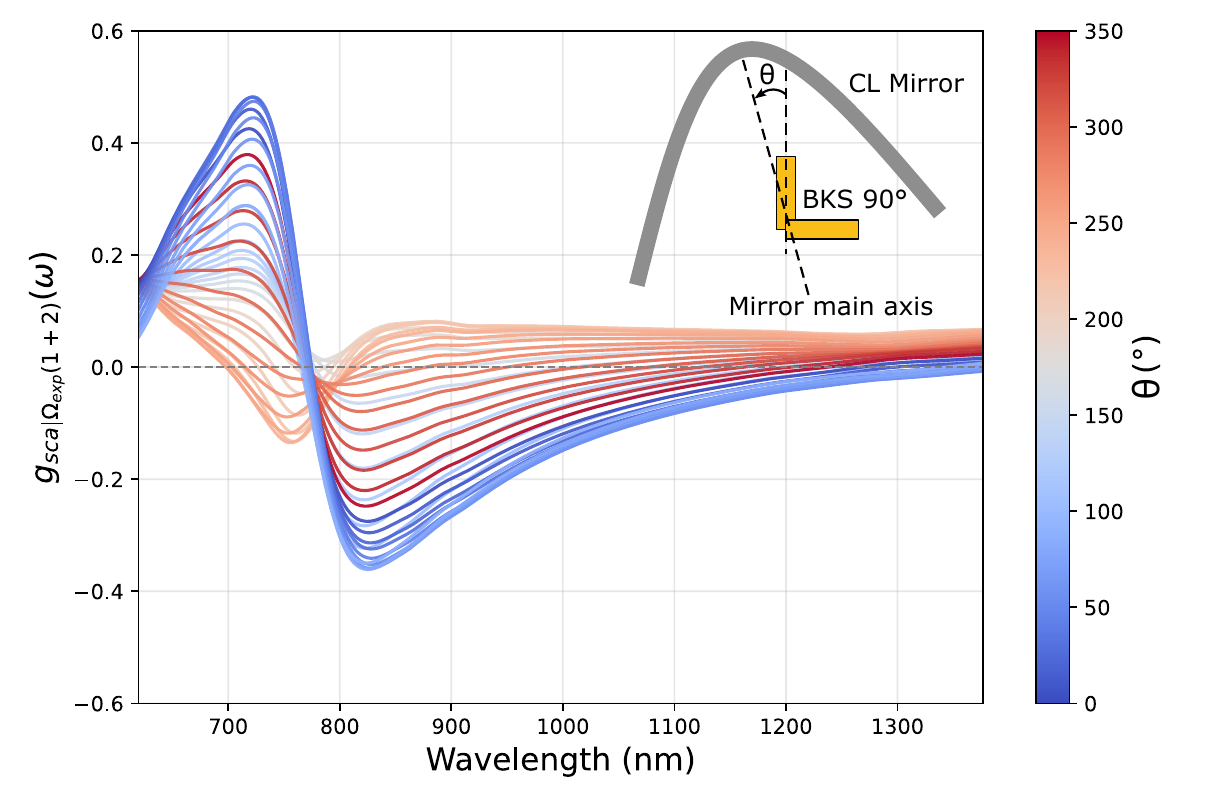}
    \caption{\textbf{Influence of the mirror's orientation on the measured dichroic signal of a 90° BKS.} Calculated pCL dichroic spectra $g(1+2)(\omega)$ of a 90° BKS collected by a real mirror (Attolight Mönch system). Here the mirror's angle $\theta$ is the angle between the bottom antenna of the BKS (vertical on all the images) and the main axis of the parabolic mirror. Experimentally mirror angle was evaluated to $\approx$ 10° which is near the optimal orientation.}
    \label{fig:Dichro90orientationmiroir}
\end{figure}

The mirror orientation is defined by the angle $\theta$ between the vertical antenna of the BKS and the main axis of the mirror, as shown in the inset of Figure \ref{fig:Dichro90orientationmiroir}. It is clear that there is a dependence between the mirror orientation and the spectrum. The optimal orientation is considered to be the one that maximizes the amplitude of the dichroic signal $g$.  


Using this mirror orientation and integrating the dichroic signals over the energy ranges of the bonding and antibonding modes, we obtain Figure \ref{fig:Dichro_Angles_BKS_B_AB}, which is the equivalent of Figure 4d when using the geometry of a real mirror. Please note that for the bonding mode and for BKS angles of 30°, 60°, 300° and 330° the simulations tend to diverge from the predictions. The exact origin of this behaviour is not known, but it is likely inherent to pyGDM2, as they do not appear when using MNPBEM.

\begin{figure}[h!]
    \centering
    \includegraphics[width=0.7\linewidth]{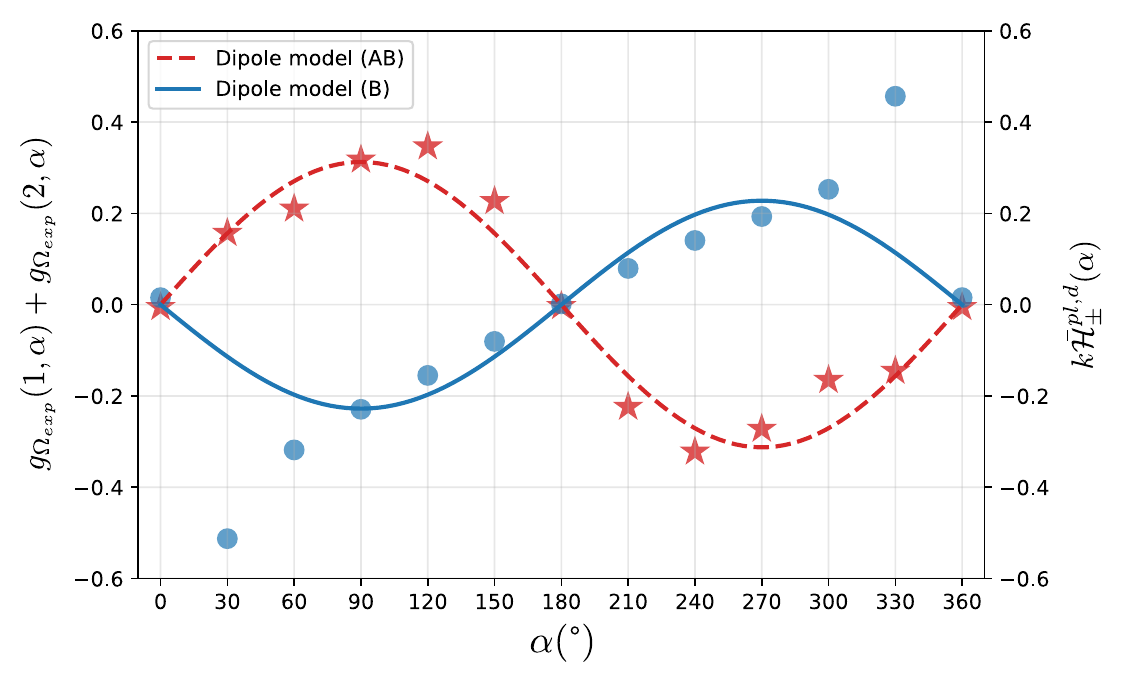}
    \caption{\textbf{Simulated dichroic signals as a function of the BKS angle $\alpha$ (pyGDM).} The dichroic signals (sum of \gscav{\Omega_{exp}} over the two tips were simulated for all kind of observed BKS for a mirror's angle set to 10° corresponding to the experimental one, and integrated on the B and AB spectral ranges.}
    \label{fig:Dichro_Angles_BKS_B_AB}
\end{figure}

\newpage

\section{Polarization modifications induced by the parabolic mirror}

It is well known that the polarization of light is affected upon reflection from a mirror \cite{Jackson:1998nia}. In the context of our study, it is crucial to determine to what extent the mirror influences the polarization measurement. The simulations presented in Figure \ref{fig:polarization_parabolic_mirror} were carried out using a custom-made Python script. For confidentiality reasons regarding the mirror geometry, the inset representation of the parabola in figure \ref{fig:polarization_parabolic_mirror} was intentionally modified.

The simulation considers an isotropic, monochromatic (800 nm) and left-circularly polarized source placed at the focus of a parabola, approximately three times larger than the actual mirror, with a silver coating. Figure \ref{fig:polarization_parabolic_mirror} shows the variation of ellipticity as a function of the reflection position on the mirror, while the inset shows the source position and the parabola geometry.

The ellipticity is defined using the $E_p = |E_p|e^{i\phi_p}$ and $E_s = |E_s|e^{i\phi_s}$ complex components, respectively perpendicular and parallel to the interface, of the reflected light by :  

\begin{equation}
    \tan(\chi) = \frac{1}{2}\arcsin{\left(\frac{2|E_pE_s|}{|E_p|^2+|E_s|^2}\sin{(\phi_p - \phi_s)}\right)}
\end{equation}

Doing so, the ellipticity is -1 (+1) for a right-handed (left-handed) polarization and 0 for a linear polarization. 

It can be observed that the collected polarization has the opposite sign to that emitted by the source. This is expected and simply results from the reversal of the propagation direction upon reflection. The Fresnel reflection coefficients were calculated for each angle and make it possible to determine the post-reflection polarization state. These coefficients depend on the composition of the interface, here vacuum–silver. For silver, the refractive index used was taken from the Johnson and Christy (1972) tables, while the refractive index of vacuum was kept constant at 1.

The solid black contour marks the region of the parabola where the polarization becomes linear ($\tan{(\chi)} =0$). On either side of this contour, elliptically polarized regions of opposite signs are observed: left-handed (L: $\tan{(\chi)} >0$) and right-handed (R: $\tan{(\chi)}<0$). The area collected by the mirror is entirely contained within the region indicated by the dashed black circle. Thus, it is clear that although the polarization is modified (mainly at the edges of the mirror), the overall signal is preserved. The average ellipticity returned by the actual mirror is -0.806. The initial polarization is therefore indeed distorted, but it does not lose its fundamental properties.

\begin{figure}[h!]
    \centering
    \includegraphics[width=0.8\linewidth]{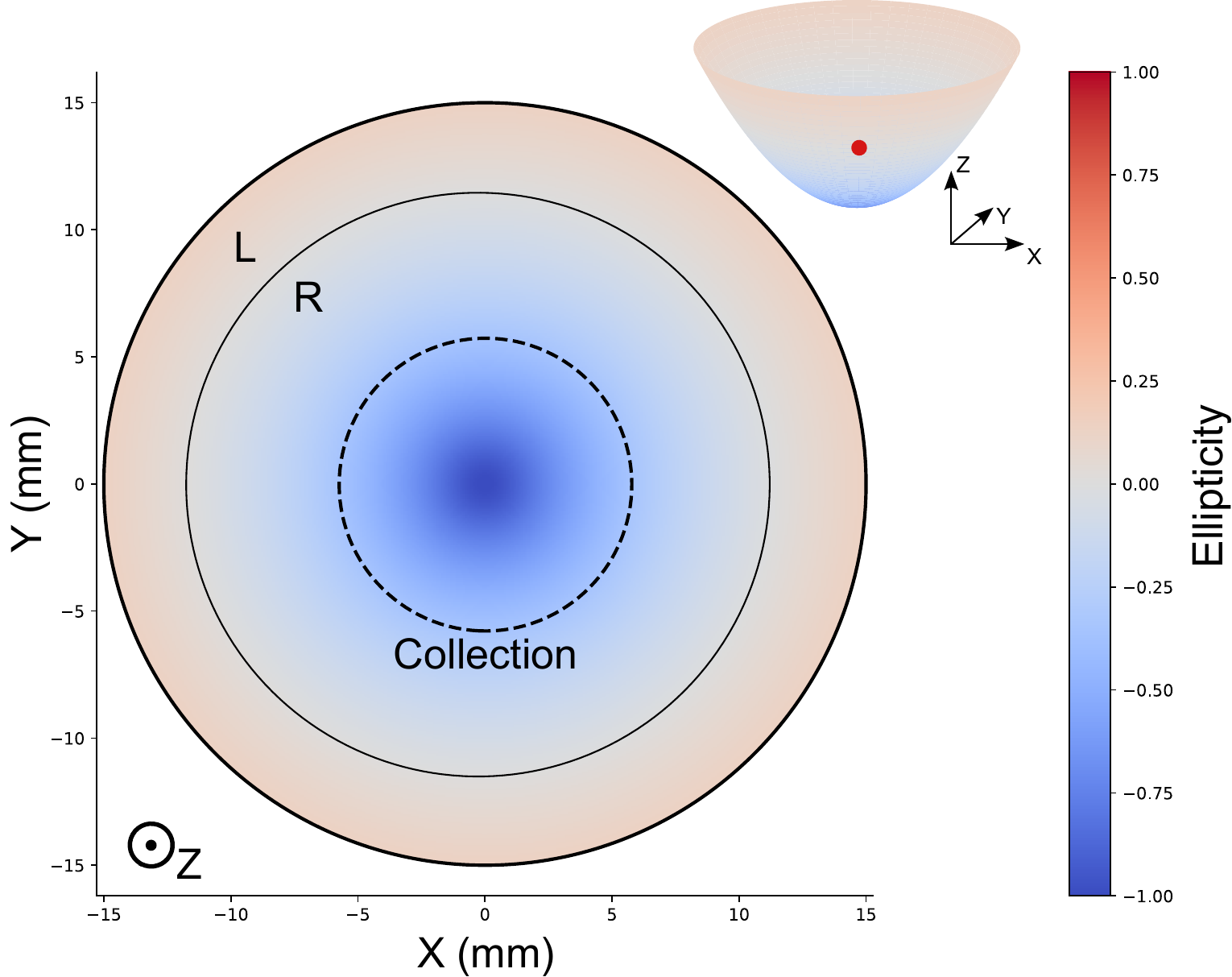}
    \caption{\textbf{Ellipticity of the signal collected with a silver-coated parabolic mirror for a monochromatic ($\lambda = 800$nm) left-handed isotropic light source located at the focal point.}}
    \label{fig:polarization_parabolic_mirror}
\end{figure}

\newpage

\section{pCL Analysis of BKS modes including higher Order Modes}

In Fig.\ref{fig:parallel pCL results}, we summarize the behaviour of the bonding, anti-bonding and higher order modes of a typical BKS ($90^{\circ}$).

As already presented in the main text, the signal of the bonding mode is mostly localized at the tips of the BKS. For the anti-bonding mode, it is mainly at the overlap, although parts of the mode are also localized at the tips. This is expected from the symmetry of the field component along the electron travel direction that depends strongly on the symmetry (bonding or antibonding) of the mode, the former being essentially null in the gap and the later strong in the gap \cite{Kociak2025}. At higher energies, however, no particular signal is observed in EELS. This is because there is no interference term in the total electromagnetic local density of state (total EMLDOS), as probed in EELS, contrary to the polarized radiative EMLDOS, probed in CL \cite{Losquin2015}.

\begin{figure}[h!]
    \centering
    \includegraphics[width = \textwidth]{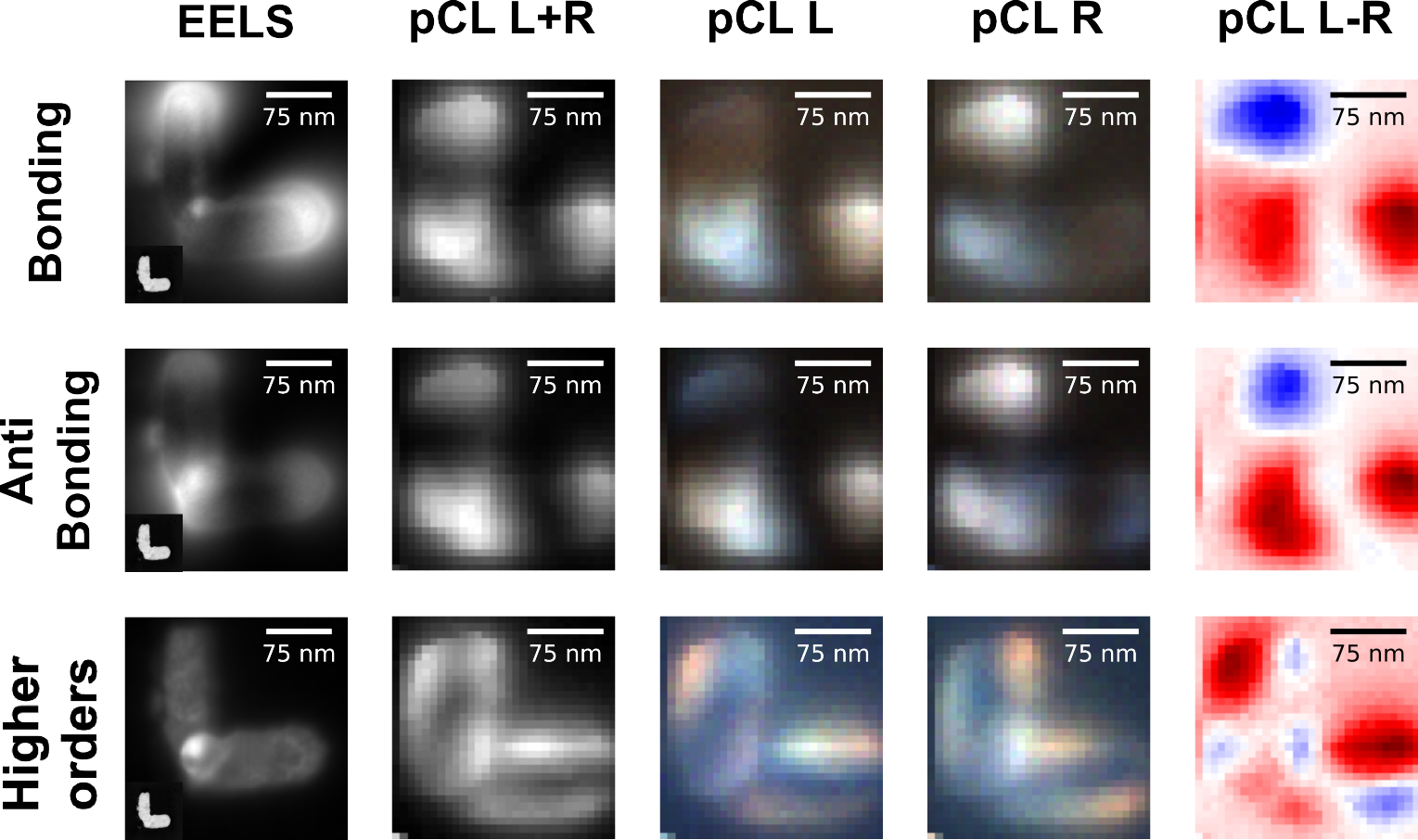}
    \caption{\textbf{Spectral mapping of EELS and parallel pCL acquisition.} From the normalized EELS (first column) and the CL (all other columns) one can compute maps filtered at different energies exemplified in the specific case of a 90$^\circ$ BKS. For the polarized left and right (pCL L and pCL R) maps, the color is coded as explained in the text and reflects the change of polarization as a function of the energy. For the sum (pCL L+R) and difference (dichroic maps, pCL L-R) maps, the signal is summed on the whole range of the considered energy windows. Energy windows from top to bottom:  bonding mode region (850 - 1070 nm), anti-bonding mode region (760 - 830 nm) and higher order modes region (530 - 640 nm).}
    \label{fig:parallel pCL results}
\end{figure}

As described earlier, our pCL setup allows us to measure two signals simultaneously: left polarization (pCL L) and right polarization (pCL R). From these signals, we can reconstruct two other signals, namely the classical CL, which is the sum of the signals (pCL L+R), and the circular dichroism, which is the the difference between them (pCL L-R).

In the second column, the pCL L+R signal is localized in the same regions as in EELS. Actually, this signal corresponds to the unpolarized CL signal of the BKS just like in Fig.\ref{fig:parallel pCL results}. However, a major difference can be observed if we perform polarimetry. Let’s start by looking at the pCL L. As can be seen for both modes, the structure emits left-handed polarized light mostly when the excited antenna is the horizontal one (the one at the top). Similarly, there is a right-handed polarized emission if the electron beam passes near the vertical antenna, which is the one embedded in the silica layer. This proves that the BKS exhibit a very local dichroic behavior. This can be calculated by subtracting the signals, thus obtaining the dichroic maps in the last column. The dichroic signal shows that while the beam is located at the tips of the antennas, the resulting emissions have opposite polarizations. However, these maps also reveal that the overlap region can also lead to polarized emissions. On these filtered maps, we also notice that the bonding and anti-bonding modes have very similar spatial distributions of polarized emissions. Nevertheless, there are some differences, particularly in the intensity of the emissions from the overlap region, which is higher for the anti-bonding mode compared to the bonding mode, which is similar to the behaviour observed in EELS. If we look at the emissions at higher energy, we observe the hidden chirality that we mentioned earlier. As it can be seen on the dichroism maps, this occurs on each individual antenna. Please note that, at those energies, there is no hidden chirality visible in the EELS map.

One can remark a singular behaviour of the spatial distribution of the hidden chirality emissions by scanning over the wavelengths. Indeed while increasing the wavelength of the displayed maps, the hotspots tend to rotate clockwise for the pCL L and counterclockwise for the pCL R. A representation of this behaviour is plotted Fig.\ref{fig:parallel pCL results} for the pCL L and the pCL R, by colorizing in blue the emissions lying between 530 nm and 560 nm, in green between 560nm - 590nm and in red the emission between 590 nm and 620 nm. Combined together, these maps form RGB images, where the color roughly indicates the wavelength of emission. On the high order maps, we can clearly see that the emission regions between 530 and 560 nm are not the same as these between 590 and 620 nm. Due to the large energy overlap between modes in this region, the phase change between modes varies continuously with energy, and filtered pCL maps with increasing energy values exhibit rotating patterns. 

\begin{figure}[h!]
    \centering
    \includegraphics[width = \textwidth]{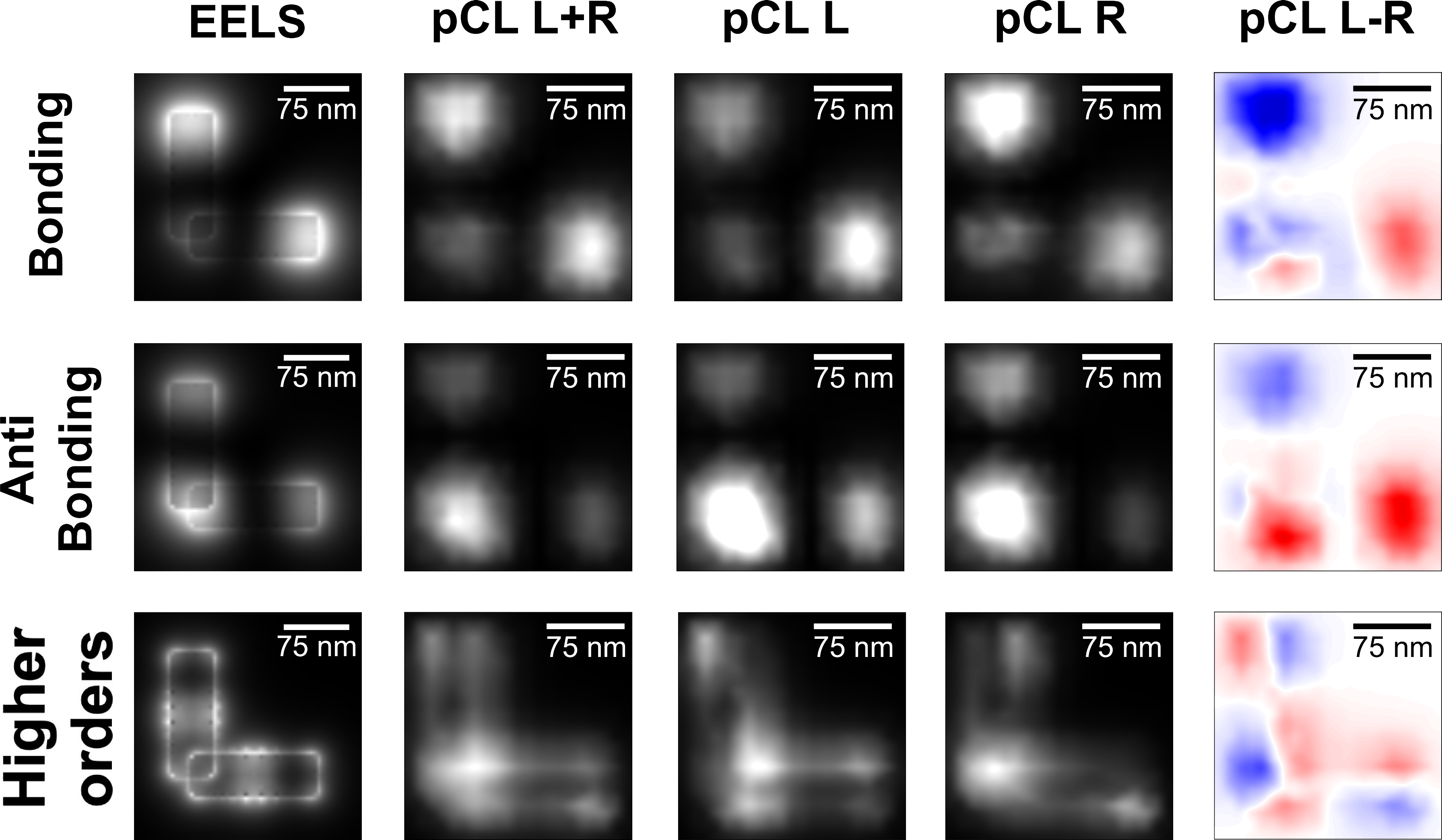}
    \caption{\textbf{Simulated spectral mapping of EELS and pCL.} EELS simulations have been performed using MNPBEM, and the pCL simulations using pyGDM2. The bonding mode has been simulated at 1.38 eV (898 nm), the anti-bonding mode at 1.68 eV (738 nm) and some higher order modes at 2.2 eV (564 nm).}
    \label{fig:simulation_parallel}
\end{figure}

Figure \ref{fig:simulation_parallel} shows simulations corresponding to the results shown in Figure \ref{fig:parallel pCL results}. The EELS simulations were performed using MNPBEM, and the pCL simulations were conducted using pyGDM2. The first notable point is the very good agreement between these simulations and the experimental results. EELS and unpolarized (Left plus Right) CL maps exhibit roughly the same behaviour. However, some differences can be seen in the pCL in the overlap region of the bonding mode. The experimental results show polarized emissions when the beam is positioned at the overlap between the two antennas, while the simulations do not. Given the spatial distribution of the bonding mode in EELS, it is likely that this hotspot is due to a spectral overlap between the bonding mode and the anti-bonding mode. Thus, part of the anti-bonding mode would be visible in the energy range of the bonding mode, which would explain the similarity between the pCL maps of these two plasmonic modes. For the higher order modes, we also observe the same behavior in pCL. The spatial distribution of the EELS is different, but this can be explained by the way the simulations were conducted. The MNPBEM maps were calculated at specific energies, which does not exactly correspond to the manner in which the experimental maps are formed, as they are integrated over a certain energy range. Therefore, when several plasmonic modes are very close spectrally, they cannot be distinguished experimentally. This is exactly what happens here: the simulations show a specific energy corresponding to a single plasmonic mode, while the experimental maps show a sum of modes.

\newpage
\section{pCL on a Single Nanoantenna}

In order to better understand the pCl behaviour of the higher order modes, we have performed pCL on isolated gold antennas (see Figure \ref{fig:schemahidden}). These antennas seem to exhibit particularly intense circular dichroism, even though these structures are achiral. This is directly the signature of the so-called "hidden chirality" effect \cite{Zu2018}, where two geometrically orthogonal modes (here the longitudinal and transverse mode) are dephased at their overlap energies. It is reminescent of the effect seen on the high order modes of the BKS.  

\begin{figure}[h!]
    \centering
    \includegraphics[width = 0.9\textwidth]{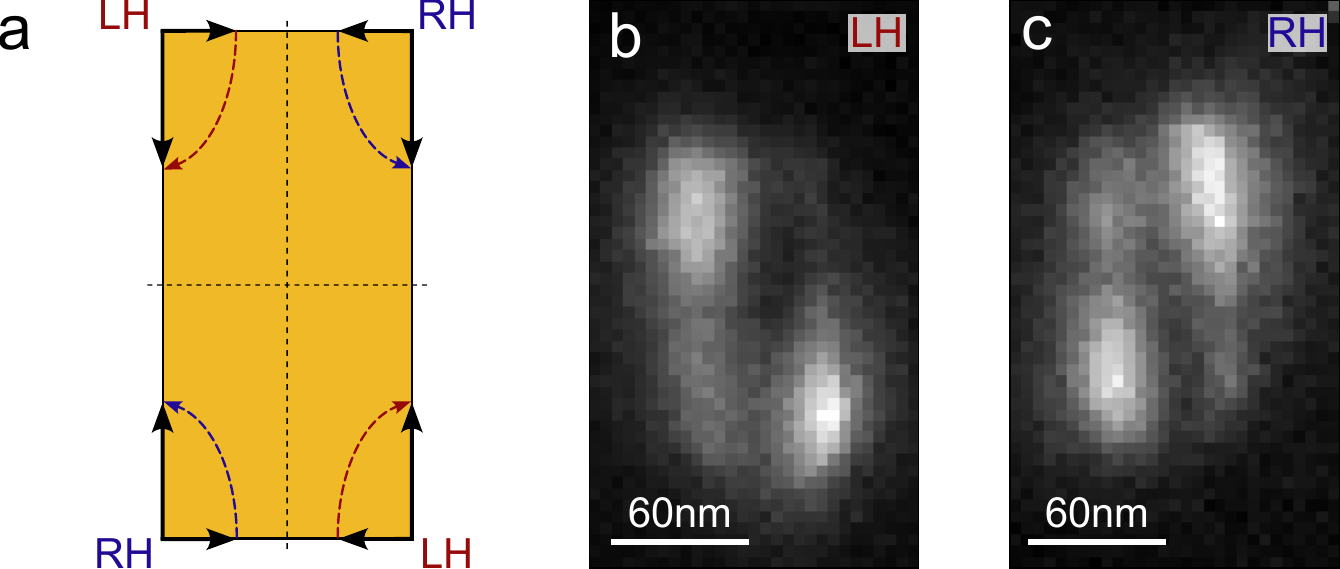}
    \caption{\textbf{Geometrical explanation of the hidden chirality on a single gold nano-antenna.} \textbf{a)} Temporal evolution of the polarization at the corners of the gold nano-rectangle. During its propagation, the light electric field needs to be aligned with both the transversal and longitudinal dipolar modes. The paths represented here are the shortest paths which enable the field to align first with the transversal mode and then with the longitdinal one. This figure is adapted from \cite{Hashiyada2014}. The dashed lines indicate the symmetry axes. \textbf{b)} Filtered left-handed (LH) pCL spectral image of hidden chirality on a gold nano-antenna. The energy range was set to 581 nm - 631 nm. \textbf{c)} Same as b) for the right-handed (RH) pCL signal. }
    \label{fig:schemahidden}
\end{figure}

\newpage

\putbib[pCL]

\end{bibunit}
\fi
\end{document}